\documentclass[aps,prb,reprint,twocolumn,nofootinbib,floats,floatfix]{revtex4-2}   

\usepackage[utf8]{inputenc}     
\usepackage[T1]{fontenc}        
\usepackage{lmodern}            

\usepackage[margin=1in]{geometry}
\usepackage{setspace}           

\usepackage{amsmath,amssymb,amsthm}
\usepackage{mathtools}

\usepackage{graphicx}           
\usepackage{booktabs}           
\usepackage{caption}
\usepackage{float}              

\newcommand{\Cpp}{C\nolinebreak[4]\hspace{-.05em}\raisebox{.25ex}{\scriptsize ++}}
\newcommand{\lr}[1]{\left(#1\right)}
\newcommand{\avg}[1]{\left\langle #1 \right\rangle}

\usepackage{hyperref}

\begin{document}

\title{Lattice in Line: Optimized DMRG ordering for complex lattice geometries}
\author{
Roman Rausch\\
\url{https://github.com/spinflip/lattice_in_line}
}
\date{\today}

\begin{abstract}
The density-matrix renormalization group (DMRG) is a one-dimensional tensor-network technique, but it is not limited to one-dimensional systems: it can be applied to periodic 2D and 3D clusters and molecules, provided their sites are first enumerated along a line; a step one may call ``lattice compilation''.

This paper discusses three proxy loss functions for finding this optimal enumeration: the graph bandwidth $B$ (maximum interaction range), the cutwidth $C$ (maximum number of bonds crossing a cut), and the average interaction range $R$. Constructing the Hamiltonian MPO (matrix-product operator) for a large set of clusters that are of interest in frustrated magnetism, I find that $C$ determines the peak SU(2) Heisenberg MPO bond dimension and $R$ the average one. Targeting $(C,R)$ lexicographically yields the best energies. Targeting $B$ indirectly reduces $C$, but not as efficiently as targeting $C$ directly. Otherwise, the value of $B$ itself is largely irrelevant in the sense that good DMRG energies can have large $B$.

To perform the optimization, classic heuristics (e.g.\ reverse Cuthill--McKee) prove unreliable even for small clusters, and I find that a QUBO formulation improves them only marginally. 
Instead I propose a staged optimization built on Boolean satisfiability (SAT) and constraint-programming (CP) solvers, CP-SAT of Google's OR-Tools (a hybrid of CP propagation and SAT clause learning) and the SAT solvers CaDiCaL and Glucose. This approach yields significantly better orderings together with rigorous bounds. 

The corresponding \textit{Lattice in Line} code is available at \url{https://github.com/spinflip/lattice_in_line} and was designed with extensive use of the Fable 5 large language model.
\end{abstract}

\maketitle

\section{Introduction: lattice compilation}

When applying the density-matrix renormalization group (DMRG) to clusters like molecules, periodic 2D and 3D lattices, we need to make them fall in line, i.e. optimally enumerate the cluster sites to fit the 1D tensor-network geometry of DMRG. This problem appears for any couplings on complex geometries. A prominent example is the Heisenberg Hamiltonian with general exchange interactions $J_{ij}$,
\begin{equation}
H
=
\sum_{ij} J_{ij}
\mathbf S_i \cdot \mathbf S_j.
\end{equation}
Another example is the Hubbard Hamiltonian with general hoppings $t_{ij}$
\begin{equation}
H
=
\sum_{ij\sigma} t_{ij}
\left(
c_{i\sigma}^{\dagger} c_{j\sigma}
+
c_{j\sigma}^{\dagger} c_{i\sigma}
\right)
+
U \sum_i n_{i\uparrow} n_{i\downarrow}.
\end{equation}

The inputs to the DMRG algorithm are the graphs $J_{ij}$ ($t_{ij}$) whose vertices we are free to enumerate. Any reordering is a unitary transformation that does not change the physics, but can improve the performance of the DMRG algorithm down the line. Because the wavefunction is expressed as a one-dimensional MPS (matrix-product state) with finite entanglement resources at each bond, the accuracy is governed by the entanglement structure induced by the chosen ordering, and not just by the physical couplings alone: Any two sites $i$ and $j$ that are adjacent on the graph can become far apart on the line, and the correlations between them have to be carried through all $\left|i-j\right|-1$ sites in between.

\begin{figure}
\includegraphics[width=0.7\columnwidth]{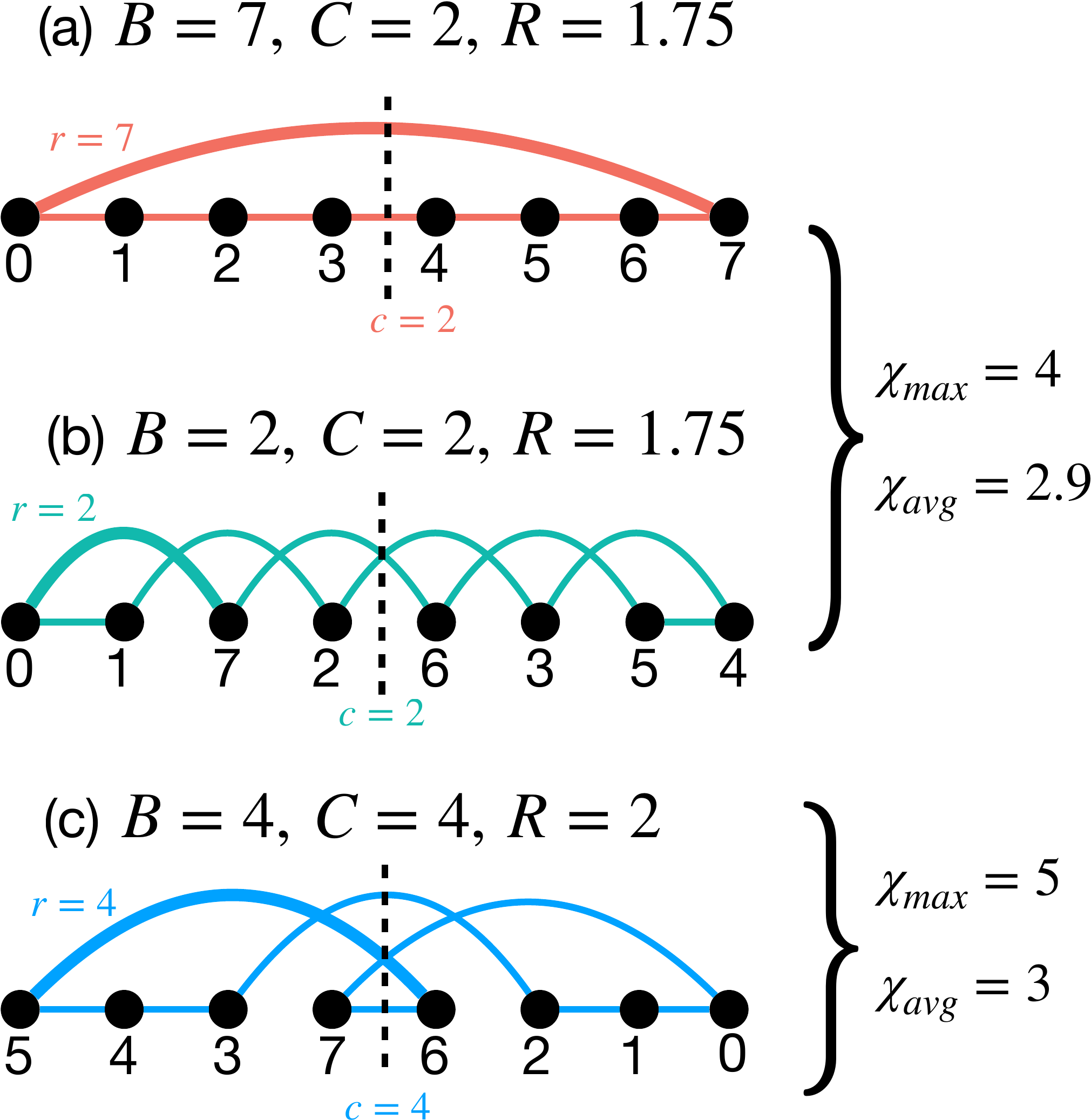}
\caption{\label{fig:ring}
Three different mappings of the periodic ring to a 1D chain. $B$ is the maximum interaction range (``bandwidth'', the corresponding longest bond is indicated by a fatter line), $C$ is maximum number of cut bonds (``cutwidth'', the corresponding thickest cut is indicated by the vertical dashed line), $R$ is the average interaction range (``mean edge length'').
}
\end{figure}

We can optimize the ordering by finding a suitable loss function to minimize. The ground truth for this problem is in principle the ground-state energy itself; in practice we need a cheap proxy before investing resources into a hard problem. Intuitively, we need to pick an enumeration that makes the graph $J_{ij}$ as close to a tridiagonal tight-binding chain as possible, where DMRG is known to be most performant.
For this, we can seek a permutation matrix $\underline{P}$ that makes the transformed matrix $\underline{P}^T\underline{J}~\underline{P}$ band-diagonal with a minimized \textit{bandwidth}. In physical terms this is the maximum \textit{interaction range}\footnote{When trying to enumerate a cluster by hand, this is the easiest quantity to keep track of.}.

In graph-theoretical terms this problem is called the \textit{linear bandwidth problem} (LBP): Given the interaction graph $G=(V,E)$ of a cluster with $L=\left|V\right|$ sites, we seek a bijective labeling $\varphi : V \to \{0,\dots,L-1\}$ minimizing $B(G,\varphi)=\max_{(u,v)\in E}\,|\varphi(u)-\varphi(v)|$. The linear bandwidth problem is NP-hard~\cite{Papadimitriou1976,ChinnSurvey1982}.

However, the bond dimension of the matrix-product operator (MPO) representation of the Hamiltonian is most closely correlated with another measure, the cutwidth $C$, which is the maximum number of bonds crossing a cut (cf. Fig.~\ref{fig:ring}). The two can be quite different: For a binary tree optimally enumerated from the top, $B$ grows faster than $C$; for a fully connected graph, $B$ grows as $L$, while $C$ grows as $L^2$. However, the two are correlated in the sense that
\begin{equation}
C \leq \sum_{k=0}^{B-1}(B-k) = B(B+1)/2.
\label{eq:ineqBC}
\end{equation}
This follows from taking any cut and summing the worst-case bonds that cross it.
In practice, this relationship implies that when minimizing $B$, one is unlikely to get a too large $C$.

Since there can be multiple solutions with minimal $B$ or $C$, we should choose a secondary objective to refine the solution. For this secondary loss function, I propose the \textit{average interaction range} $R$ (cf. Fig.~\ref{fig:ring}), which is directly related to the \textit{average} MPO bond dimension.

In this paper, I discuss the role of $B$, $C$ and $R$ for this process of ``DMRG lattice compilation'' and will argue that optimizing for $C$ first and for $R$ second is the best approach, while the value of $B$ remains largely irrelevant. I also provide a SAT-based code to compute this enumeration with rigorous bounds. The code is meant to help DMRG practitioners and is largely generated by Fable 5 as a private project.

I note that the lattice compilation problem bears close similarity to the routing stage of quantum circuit compilation\footnote{Usually called ``transpilation'' for historical reasons.}~\cite{Wille2019,Tan2020}, where SAT-based codes have recently been proposed~\cite{Molavi2022,Lin2023,Shaik2023,Shaik2024,Yang2024,Jakobsen2025,Milkevych2026}.

Note that the graph input for lattice models is simpler than in quantum chemistry, where one has dense all-to-all hoppings and interactions, and where the same enumeration problem appears. There, genetic algorithms were introduced to tackle it early on, first based on the energy alone~\cite{Moritz2005DMRGOrdering}, and later using cheaper proxies based on distance and orbital overlap~\cite{Zhai2023Block2}. They are tailored to long-range interactions for fermions, where the graphs are dense and are beyond the scope of this discussion.

\section{\label{sec:lattice}Exploiting lattice translations does not help}

One might think that for structured graphs like periodic lattices with a unit cell, the problem can be separated into a minimization of $B$ (resp. $C$) of the intercell graph, followed by an optimized enumeration of the sites within the unit cell. In this way, the problem can be broken down into two smaller ones, dealing only with $N_{\text{cells}} = L / L_{\text{cell}}$ vertices in the first step. However, this approach is probably only optimal in the asymptotic limit of large lattices $N_{\text{cells}} \gg L_{\text{cell}}$. For typical lattice problems treated in DMRG, we have only small clusters with a couple ($O(1)$-$O(10)$) of unit cells in the periodic direction. A better solution in this case is usually given by a nontrivial interleaving of the unit cell sites.

In the case of cylinders (periodic boundary conditions (PBC) in y-direction, but not in x-direction) this interleaving can be guessed because the solution is broadly given by the infamous ``snake path''\footnote{For the hellenophile, a \textit{boustrophedon}.} running left-to-right and winding across the cylinder~\cite{Stoudenmire2012}, and one can typically find an optimal path through the unit cell by hand~\cite{Depenbrock2012Kagome}.
However, the problem becomes genuinely nontrivial for 2D and 3D tori with full PBC, as well as for generic molecules, so that a more systematic approach that includes the snake as a particular solution is needed.

\section{\label{sec:heuristic}Heuristic Methods}

Various heuristic algorithms have been formulated for graph linearization since the 1960s. For an overview and implementation, see Ref.~\cite{Varona2025MatrixBandwidth}. Relevant examples are:
\begin{itemize}
\item Reverse Cuthill--McKee (RCM)~\cite{cuthill1969reducing,george1971computer}
\item Gibbs--Poole--Stockmeyer (GPS)~\cite{gibbs1976algorithm}
\item Sloan~\cite{Sloan1986ProfileWavefront}
\item King~\cite{King1970AutomaticReordering}
\item Spectral (Barnard--Pothen--Simon) Ordering~\cite{Barnard1995SpectralEnvelope}
\end{itemize}
These algorithms are designed to minimize different metrics: RCM targets $B$, the ``reverse'' algorithm improves over the regular one on the ``profile''. GPS targets $B$ as well. Sloan and King target the ``wavefront'', which is more closely related to $C$ and $R$. Spectral ordering targets the ``total edge length'', which is related to $R$. 

A benchmark of the influence of these algorithms on $B$, $C$ and $R$ for DMRG geometries of interest can be found in Tab.~\ref{tab:heuristic} of App.~\ref{app:heuristic-benchmark}. In general we can say that RCM is indeed best for $B$, Sloan and King are best for $R$, while the spectral ordering turns out to be best for $C$. 

RCM was included into DMRG codes as early as 2002-2005~\cite{chan2002highly,Moritz2005DMRGOrdering}. 
A benchmarking of different heuristic orderings for the icosidodecahedron ($L=30$) Heisenberg molecule~\cite{ummethum2013large} showed comparable downstream energy performance of RCM, Sloan, and a hand-picked enumeration\footnote{It turns out to have optimal $C=12$ for this problem.} as opposed to a random enumeration.

\begin{figure}
\includegraphics[width=0.7\columnwidth]{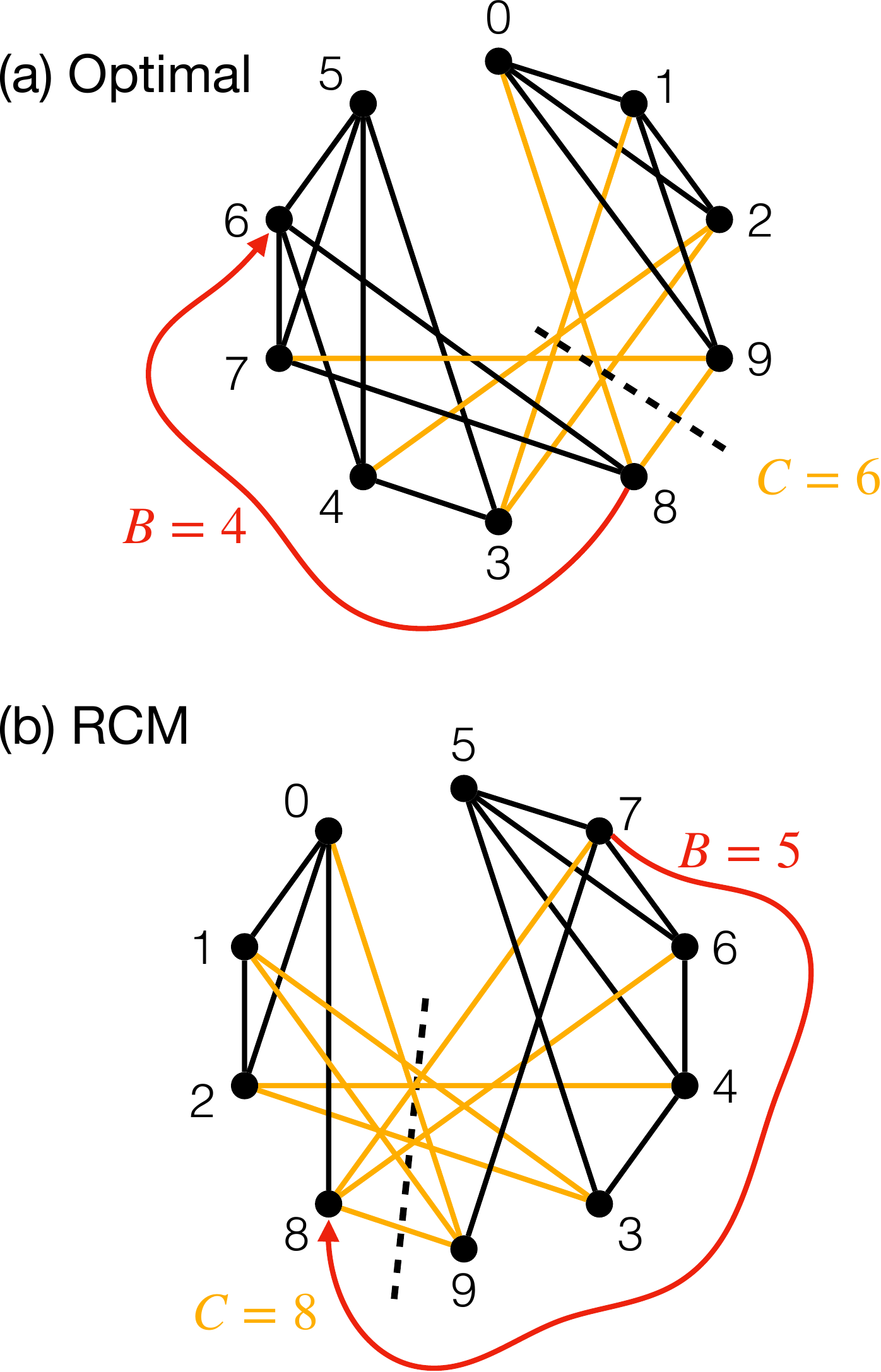}
\caption{\label{fig:ring2}
Shortcoming of the Reverse Cuthill--McKee (RCM) algorithm for a ring with $L=10$ and $J=J'=1$: (a) The solution, optimal in both bandwidth $B$ and cutwidth $C$; (b) the RCM solution.
The resulting chain is bent to a horseshoe for better visibility and the numbers indicate the sites in the original enumeration to verify connectivity.
The orange bonds indicate the position with the largest number of cuts; the red arrow indicates the longest hopping in the resulting chain.
}
\end{figure}

The heuristic algorithms can, however, fail to optimize either $B$ or $C$. For example, for the triangular and kagome cylinders, the algorithms surprisingly tend to worsen the venerable snake path (cf. Tab.~\ref{tab:heuristic}). Another curious case is the one of the $J-J'$ periodic chain (denoting nearest- and next-nearest-neighbour interaction; for simplicity we set $J'=J=1$). For $L=10$, there is one solution that optimizes both $B$ and $C$ with $B=4, C=6$, see Fig.~\ref{fig:ring2}. However, RCM converges to $B=5, C=8$. Hence, RCM is probably not a good method of choice for chains and rings with a unit cell, at least not without cross-checking\footnote{Note also that RCM can sometimes yield slightly different results based on the implementation. Here, I use the SciPy~\cite{Virtanen2020} backend \texttt{scipy.sparse.csgraph.reverse\_cuthill\_mckee}.}. RCM's first heuristic is to sort the vertices by degree, and that gives no meaningful ranking in the presence of translational symmetry.

Heuristic algorithms should be regarded as cheap baselines that only provide the upper limits on $B$ and $C$. For small or 1D-like problems we can usually get away with a suboptimal enumeration by just scaling up the bond dimension, but more complex geometries require going beyond.

\begin{figure*}
\includegraphics[width=0.9\textwidth]{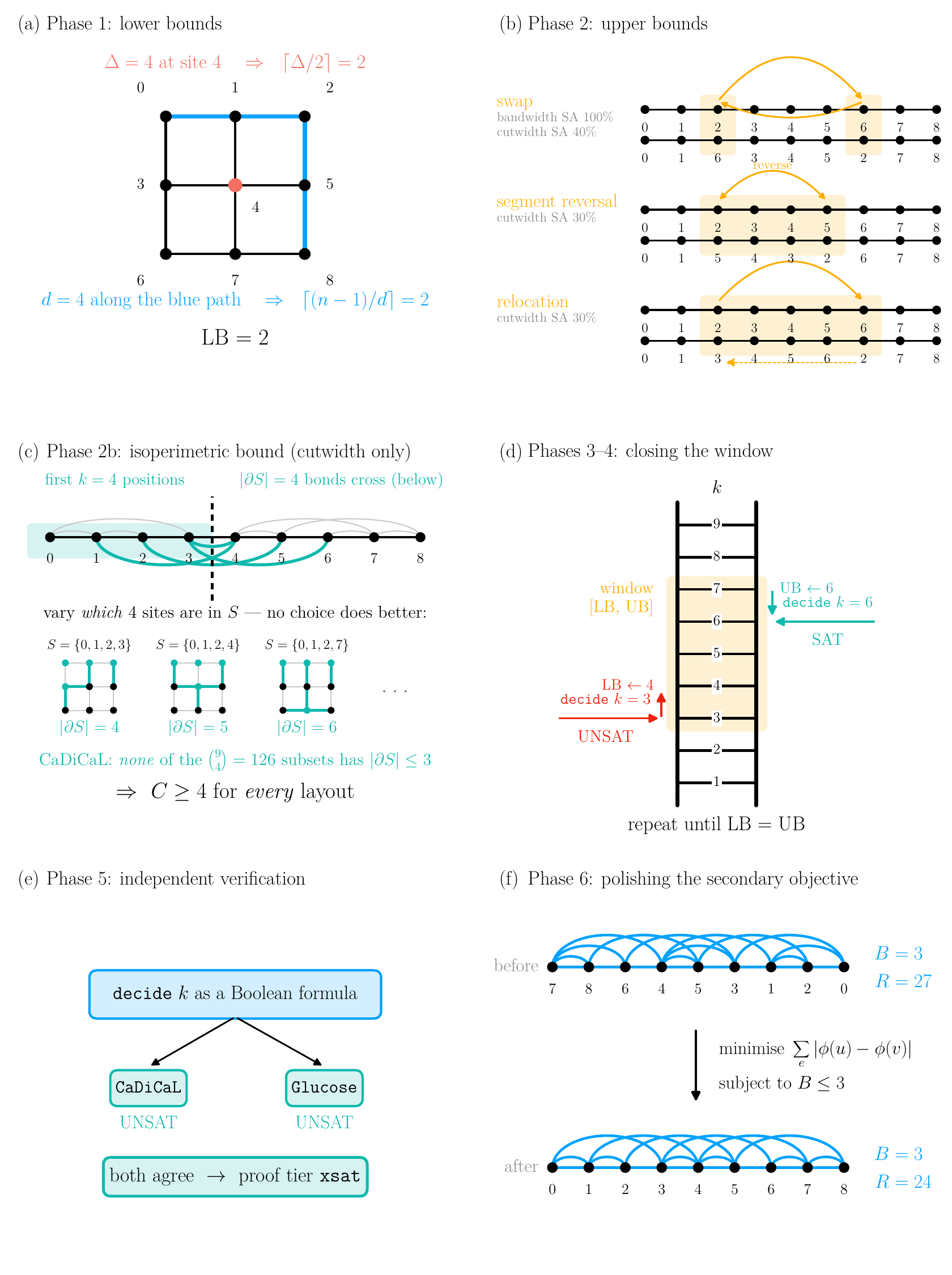}
\caption{\label{fig:certify_phase_detail}
Visual guide to the phases in the SAT campaign: 
(a) analytical bounds (only diameter bound and degree bound shown; cf. Fig.~\ref{fig:certify_congestion} for the congestion bound).
(b) lowering the upper bound by simulated annealing.
(c) using SAT to improve the lower bound for the cutwidth via the isoperimetric bound.
(d) decision ladder: trying to prove satisfiability (SAT) from above and unsatisfiability (UNSAT) from below.
(e) independent UNSAT verification from CaDiCaL and Glucose.
(f) polishing phase to reduce the secondary objective of the average interaction range $R$.
}
\end{figure*}

\section{\label{sec:qubo}QUBO gridlocked}

To go beyond heuristic algorithms, we need to deal with the full combinatorial problem. One proposition is to map the linear bandwidth problem to a QUBO (quadratic unconstrained binary optimization = Ising glass)~\cite{GuoDinneen2025}. This is motivated by exploring how to solve these problems on a quantum computer, but in fact we have powerful classical QUBO solvers that vastly outperform current quantum computers. By using tabu search~\cite{glover1989tabu,glover1997tabu,Liang2022,Liang_libtsqubo}, I was unable to obtain a meaningful improvement in $B$ with the QUBO approach except for cylinders (cf. Tab.~\ref{tab:bandwidth}). Moreover, there is no easy way of also integrating the secondary objective $R$ into the QUBO technique. I leave the code in an experimental stage for anyone else to try. More details about the mapping are discussed in App.~\ref{app:qubo}.

\section{\label{sec:sat}Breaking bad bonds with SAT}

A very powerful and systematic approach to the problem is by way of the Boolean satisfiability problem (SAT). SAT asks whether a logical formula in Boolean variables can be satisfied. It is a hard problem for which highly optimized solvers nonetheless exist. The question ``Is there a site ordering of bandwidth (cutwidth) at most $B$ ($C$)?'' maps directly onto such a formula: Boolean variables assign labels to sites, and clauses enforce a valid ordering in which no coupled pair is placed more than $B$ apart. This approach gives algorithms more information to work with than just QUBO energies. We can sweep the values of $B$/$C$, trying to answer the above question each time, which provides rigorous bounds; and we can even certify the optimal solution if the bracket is closed.

Note that a QUBO problem can always be converted into a SAT problem, but the QUBO formulation here is unnatural because the problem is non-quadratic, and the resulting QUBO becomes bloated. Instead, we can apply SAT in a more native fashion.

The provided code (``Lattice in Line'') returns bounds on $B$ (resp. $C$), and is based on approaches to the related cyclic bandwidth problem in constraint programming~\cite{CodognetMonfroy2025,GuoDinneen2025}. It uses multiple phases, and is thus a full ``campaign'' designed to systematically hack away at the problem from various directions, and not just a single-shot attempt. The steps of the campaign are described in the next section, a visual guide to follow is provided in Fig.~\ref{fig:certify_phase_detail}.

\subsection{The SAT campaign}

\begin{figure*}
\includegraphics[width=\textwidth]{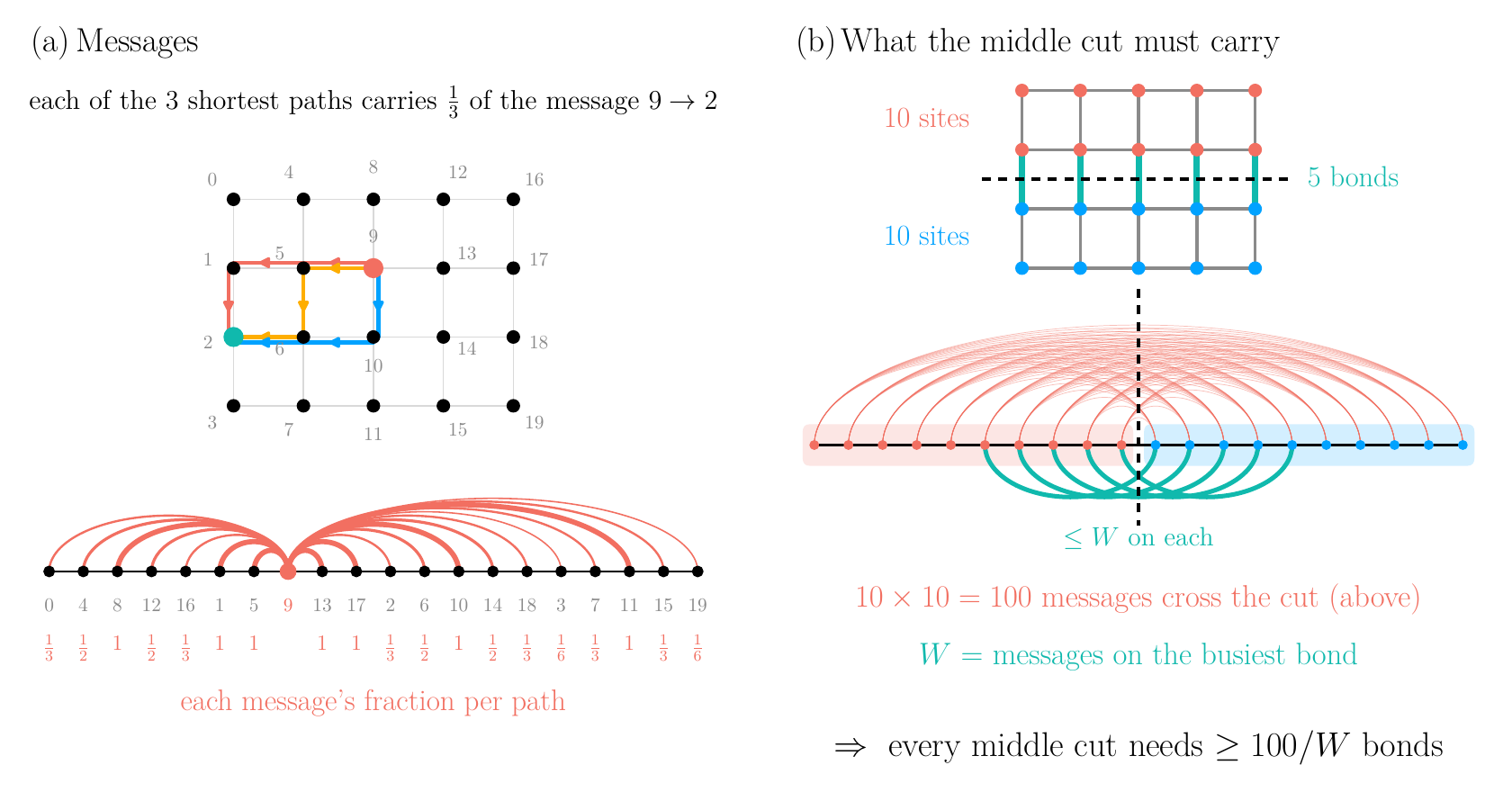}
\caption{\label{fig:certify_congestion}
Visual guide to the congestion approach:
(a) definition of a ``message''.
(b) definition of $W$, the load on the busiest bond.
}
\end{figure*}

\paragraph{Phase 1: lower bounds [Fig.~\ref{fig:certify_phase_detail} (a)].}
The lower bound $\mathrm{LB}$ is set as the maximum of several cheaply verifiable combinatorial quantities known from graph theory. For $B$, this is the degree bound, the diameter bound, and a subgraph/ball refinement. For $C$, this is max-degree, degree-sum and spectral bound.

Another bound used here is the congestion bound~\cite{Leighton1999}, which is less well known and deserves an explanation. Imagine that every pair of sites exchanges one message, which travels along the bonds of the lattice on a shortest path between the two; if there are several equally short paths, the message is split evenly among them (cf. Fig.~\ref{fig:certify_congestion} (a)). For each bond we add up the (possibly fractional) messages passing through it, and call $W$ the load on the busiest bond (cf. Fig.~\ref{fig:certify_congestion} (b)). Now take any ordering and cut the chain in the middle. There are $L/2$ sites on either side, so $(L/2)^2$ pairs of sites are separated by the cut, and the message of each such pair has to cross it on one of the cut bonds. Since no bond carries more than $W$ messages, the cut must contain at least
\begin{equation}
C \;\ge\; \frac{(L/2)^2}{W}
\label{eq:congestion}
\end{equation}
bonds~\footnote{For example, on the $10\times10$ square torus $50\times50=2500$ messages must cross the middle cut and the busiest bond carries $W=125$ of them, so that $C\ge 20$ for any ordering (cf. the smaller rectangular plaquette in Fig.~\ref{fig:certify_congestion}).}.

\paragraph{Phase 2: upper bounds [Fig.~\ref{fig:certify_phase_detail} (b)].}
The upper bound $\mathrm{UB}$ is first set by the quick RCM algorithm for $B$. From here, the code tries to improve on it by CPU-parallel multi-start simulated annealing runs over transpositions of $\varphi$ (50\% are random swaps, 50\% target to shorten the longest bond). For $C$, the code omits the RCM stage and does the following operations in the annealing step: 40\% pair swaps (pick two random vertices and transpose), 30\% segment reversal (pick two random locations and reverse the whole slice), 30\% relocation (pick a random vertex and insert it at another random location, causing a cyclic shift of the intervening block).
In both cases, the secondary objective $R$ is included lexicographically, i.e. one always optimizes over $B$ (resp.\ $C$) first, and over $R$ second. The jobs stop early on stagnation. This phase does the bulk of the work to actually get a good solution.

\paragraph{Phase 2b: isoperimetric bound ($C$ only) [Fig.~\ref{fig:certify_phase_detail} (c)].}
Cut the chain at the first $n$ positions. These $n$ sites form a subset $S_L$, and the bonds crossing the cut are exactly the bonds that connect $S_L$ to the rest, i.e.\ its boundary $\partial S_L$. Whatever the ordering, $\left|\partial S_L\right|$ cannot be smaller than the smallest boundary among \emph{all} subsets of $n$ sites. Hence
\begin{equation}
C \;\ge\; \max_{n}\; \min_{|S|=n} \left|\partial S\right| ,
\label{eq:isoperimetric}
\end{equation}
i.e.\ the cutwidth is at least the smallest ``surface'' that a region of $n$ sites can have, for every $n$. Finding this smallest surface is known as the \textit{edge-isoperimetric problem}, which has been solved analytically for simple lattices such as grids~\cite{Bollobas1991}. The question ``Is there a subset of $n$ sites with at most $T$ boundary bonds?'' is a much smaller SAT problem than the ordering problem itself. If the answer is ``unsatisfiable'' (UNSAT), then $C \ge T+1$.

Each such question, for one pair $(n,T)$, is answered by CaDiCaL~\cite{biere2020kissat}, a state-of-the-art open-source SAT solver. An UNSAT verdict is a strong claim, namely that none of the exponentially many subsets works, and it rests on the correctness of the solver. To guard against a software bug, every UNSAT that is produced is therefore solved a second time by the independent SAT solver Glucose~\cite{Glucose2009}. If both agree, the bound is recorded as \textit{xsat} (``cross-sat''); if Glucose does not finish within a preset time limit, it stands at the weaker level \textit{sat} (one solver only). In addition, Glucose can write out a DRAT proof: a step-by-step log of its logical deductions, which a separate checker program could replay to confirm the UNSAT without trusting either solver. The proof can optionally be saved, but the replay is omitted in practice, since it would require the additional package drat-trim~\cite{wetzler2014drattrim} (written in C), and is generally a bit of overkill for the physical problem we ultimately care about.

The code can walk through the thresholds $T$ in two directions. For small clusters it steps $T$ down from $\mathrm{UB}-1$: an UNSAT at the very first step already proves the phase-2 layout optimal, and each further step down settles for a weaker bound. For large clusters the window $[\mathrm{LB},\mathrm{UB}]$ is wide and the questions near $\mathrm{UB}$ are out of reach, so that stepping down would spend the full time limit on each of them. Here the code instead climbs up from $T=\mathrm{LB}$: every UNSAT raises the lower bound by one, and the climb stops at the first threshold for which no UNSAT is found. In both directions, each threshold is tried for all subset sizes $L/5\le n\le L/2$ in parallel (larger $n$ are covered by exchanging the subset with its complement), with a time limit per question.

The ladder can end without a proof for two different reasons. Either the questions run into the time limit, in which case more computing time may still raise the bound. Or the solver finds, for every size $n$, an actual subset with at most $T$ boundary bonds. Then Eq.~\eqref{eq:isoperimetric} itself is exhausted: regions with such a small surface exist for every size, and whether they can be lined up into a single ordering is a question this bound cannot answer, no matter how much computing time is spent.

One technical detail, found by experimentation, turns out to matter a lot. The order in which the bonds are listed in the SAT formula does not change the answer, but changes the solving time by orders of magnitude: the solver is fast when bonds that are close on the lattice are also listed close to each other, and can fail to finish within the time limit for a random order. No single order seems to be best for all lattices, so each question is posed twice. The first order lists the bonds as they appear along the phase-2 layout. The second keeps the bonds of each elementary unit (triangle or tetrahedron) together and walks from unit to unit through shared sites. An UNSAT from either one is sufficient.

Overall, phase 2b does the bulk of the work to increase the lower bound for $C$ and in fact closes the interval in many cases.

\paragraph{Phase 3: direct minimization ($B$ only) [Fig.~\ref{fig:certify_phase_detail} (d)].}
Phase 3 hands the problem to a constraint-programming solver (CP-SAT) and asks it to minimize $B$ directly, warm-started from phase 2. Since closing the problem may not be achievable on hard clusters, a timeout is imposed here as well. Within this time the solver both lowers UB and raises LB, proving bounds. Typically, this phase leaves with a narrowed interval $[\mathrm{LB},\mathrm{UB}]$ that is not closed yet. For $C$, this phase is omitted: certifying $C$ in this way is in practice far harder for CP-SAT than certifying $B$ on the lattice graphs considered here (phase 2b is designed to address it specifically), so the pass would spend its budget without improving the bounds.

\paragraph{Phase 4: decision ladder [Fig.~\ref{fig:certify_phase_detail} (d)].}
Now, the code attempts to settle the values $k\in [\mathrm{LB}, \mathrm{UB}]$ by solving $\mathrm{sat}(G,k)$ as a Boolean/constraint feasibility problem. A satisfying assignment is an explicit labeling with $B\le k$ (resp. $C\le k$), while an UNSAT verdict is a proof that none exists.

For $B$, parallel jobs are run from both ends: jobs at $k=\mathrm{UB}-1,\mathrm{UB}-2,\dots$ (each SAT result lowers $\mathrm{UB}$ to the achieved bandwidth) and jobs at $k=\mathrm{LB},\mathrm{LB}+1,\dots$ (each UNSAT result raises $\mathrm{LB}$ to $k+1$). The ladder terminates when $\mathrm{LB}=\mathrm{UB}=B^\star$. The decisive instance is the UNSAT at $k=B^\star-1$: its impossibility proof is what makes the result certified rather than just best-found. 

For $C$, the ladder is sequential rather than parallel: it first works the SAT side at $k=\mathrm{UB}-1$, warm-started from the phase-2 layout, and once that side times out, it is retired and the remaining budget goes to the UNSAT side at $k=\mathrm{LB}$. The rationale is that lowering UB gets progressively harder for $C$, and the budget is better spent increasing LB.

For hard clusters, closing the interval is not practically achievable in a reasonable runtime, and we are therefore still left with a final interval once the budget is exhausted. For $C$, a global time limit is again imposed. For $B$, the ladder instead aborts early once a round yields no verdict (every individual decision timing out).

\paragraph{Phase 5: independent verification [Fig.~\ref{fig:certify_phase_detail} (e)].}
The UNSAT verdicts of phases 3 and 4 come from CP-SAT alone (level \textit{cpsat}). Optionally, the decisive one at $k=B^\star-1$ (resp.\ $C^\star-1$) can be cross-checked by the same two solvers as in phase 2b: the ordering problem is translated into a pure SAT formula and given to CaDiCaL and Glucose concurrently. If both conclude UNSAT, the certification is raised to \textit{xsat} and the DRAT proof can be saved; if they time out, it stands at \textit{cpsat}. If the window was already closed by phase 2b, this phase is skipped.

\paragraph{Phase 6: polishing the secondary objective [Fig.~\ref{fig:certify_phase_detail} (f)].}
Finally, the code minimizes the total interaction range $\sum_{(u,v)\in E}|\varphi(u)-\varphi(v)|$ (an integer formulation of the minimization of the average interaction range $R$) among all orderings with fixed $B$ (or $C$) held at its best-known value, again via CP-SAT.

\paragraph{Optional phase: improved congestion bound ($C$ only).}
Improvement of the congestion bound is an approach that only worked in one case for the clusters considered here (see Sec.~\ref{sec:certif}), so its description is put into App.~\ref{app:opt-phase}. In the code, it can be called as a standalone phase.

The code uses the Python libraries ortools~\cite{ortools} for phases 3, 4 and 6 (CP-SAT); and python-sat~\cite{PySAT2018} for phases 2b and 5. The optional phase uses HiGHS~\cite{Huangfu2018} through SciPy~\cite{Virtanen2020}.
Optionally, pynauty~\cite{pynauty,nauty2014} computes the vertex orbits of the graph. In the ordering problems (phases 3--6), the vertex receiving the first index is restricted to one representative per orbit, which reduces the search space by the orbit size $L/N_{\mathrm{orb}}$ (for orbits of equal size); without pynauty, a reversal-symmetry constraint is used instead (a factor of 2). In phase 2b, one site is fixed inside the subset if all sites are symmetry-equivalent ($N_{\mathrm{orb}}=1$); otherwise this is only possible at $n=L/2$. Obviously, using pynauty is highly recommended.

\section{Results}

\subsection{Choice of clusters}

The clusters chosen are mainly frustrated systems that are of high interest in the quantum theory of magnetism. They fall into the following categories:
\begin{itemize}
\item \textbf{Archimedean solids}: The truncated tetrahedron is an easy baseline, the icosidodecahedron is a well-studied Heisenberg molecule~\cite{ummethum2013large}, the C$_{60}$ is the famous buckminsterfullerene~\cite{Rausch2021C60,rausch2026pair}.
\item \textbf{Other molecules}: Only the fullerenes C$_{20}$ and C$_{40}$ are included; other nanoclusters like triangulenes, coronenes and larger fullerenes are in principle of interest.
\item \textbf{Kagome \& Triangular}: well-known challenging 2D lattices, both cylinders and tori with up to $O\lr{10^2}$ sites (a size which is still DMRG-solvable for the Heisenberg model).
\item \textbf{Hyperkagome}: a frustrated 3D lattice; the most interesting cluster has $L=324$ because this is the smallest one that reproduces the loop-10 localized magnons of the bulk lattice~\cite{Hutak2024}.
\item \textbf{Pyrochlore \& Trillium}: frustrated 3D lattices; the pyrochlore clusters considered here are identical to Ref.~\cite{hagymasi2021possible}.
\item \textbf{Square torus}: not frustrated, but added into the mix as a fundamental baseline.
\end{itemize}

\begin{figure*}
\includegraphics[width=0.7\textwidth]{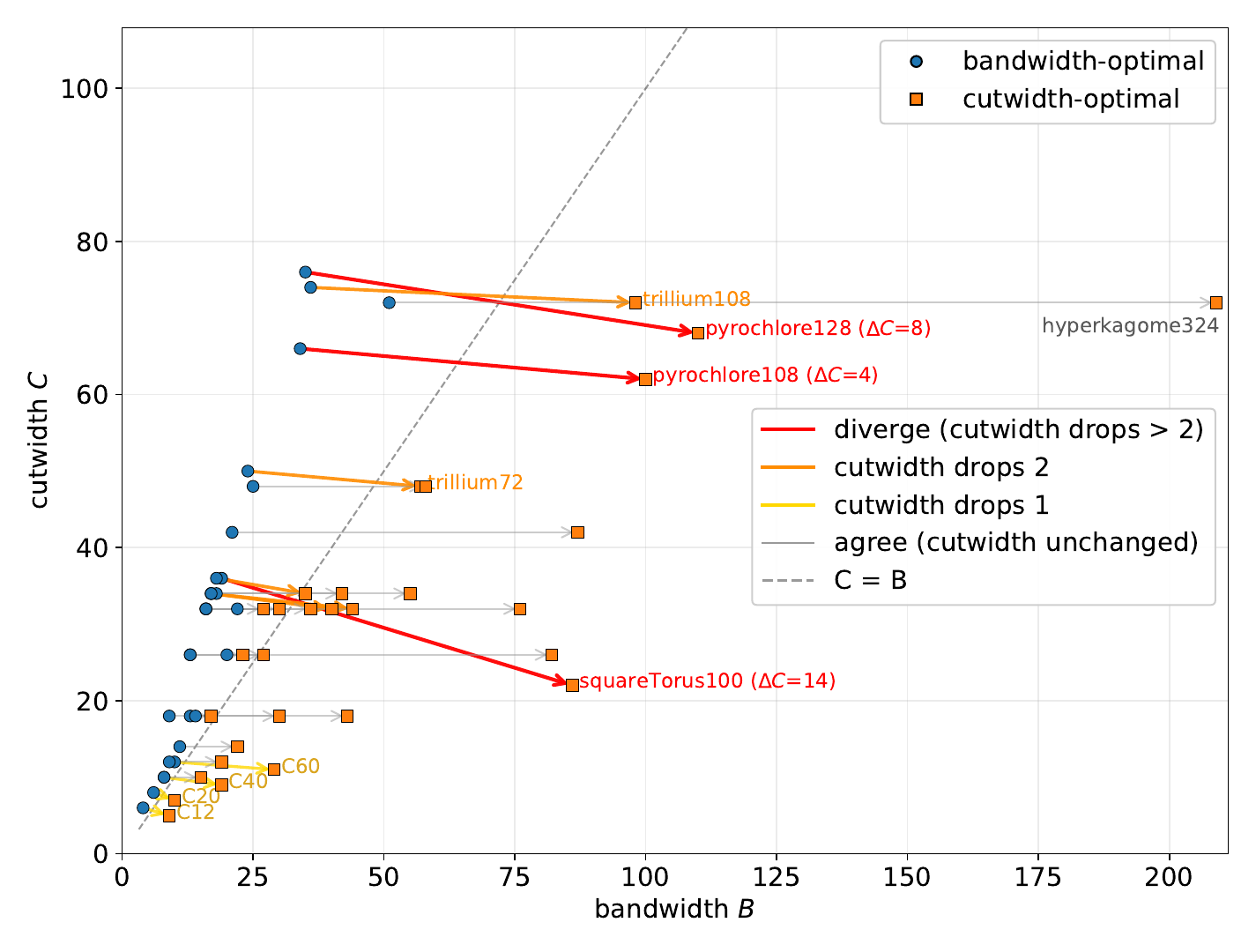}
\caption{\label{fig:mpo_divergence}
Scatterplot of the obtained campaign solutions (Tabs.~\ref{tab:bandwidth} and~\ref{tab:cutwidth}) in the cutwidth $C$ vs. bandwidth $B$ plane. When the $B$ and $C$ campaigns led to the same $C$, the points are joined by a grey line, differences of 1 (2) are marked by a yellow (orange) line; and larger differences are marked by a red line.
}
\end{figure*}

\begin{figure}
\includegraphics[width=\columnwidth]{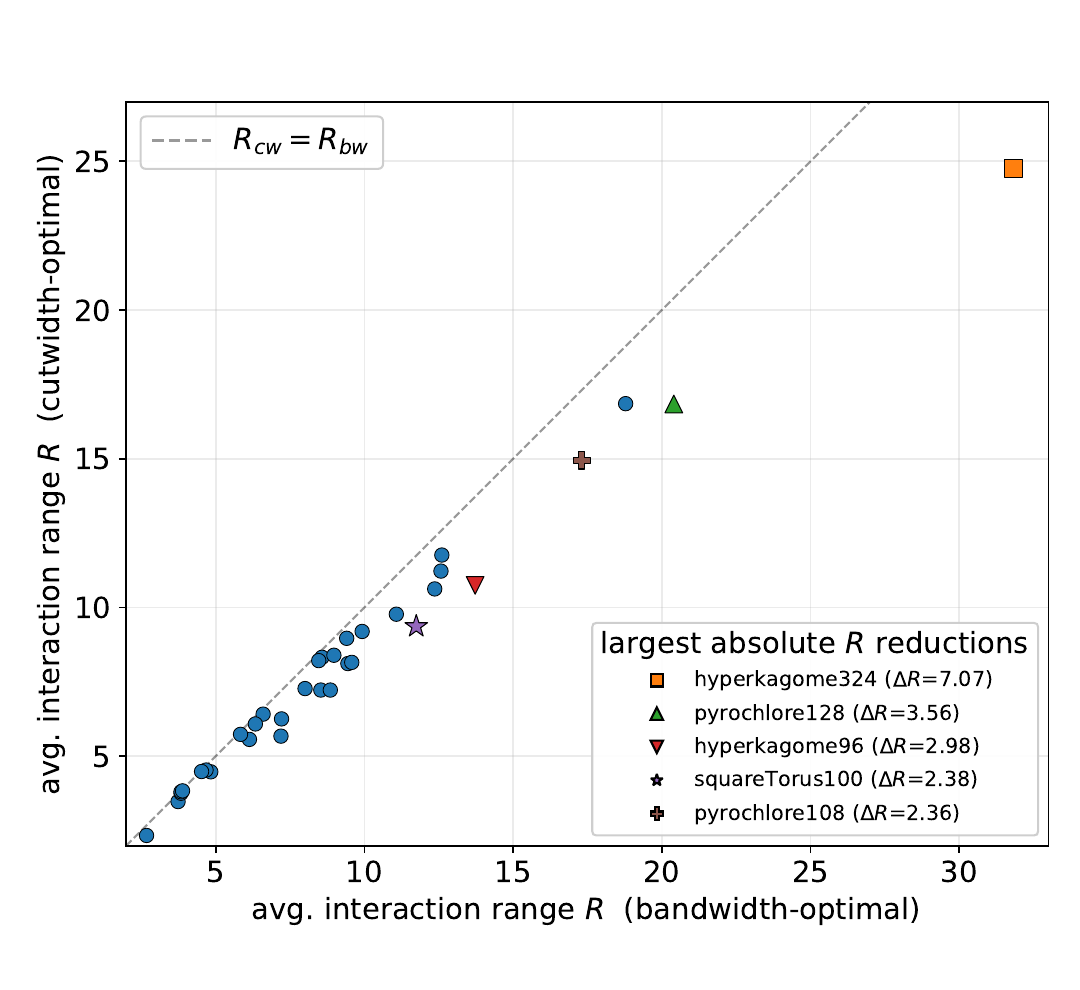}
\caption{\label{fig:mpo_avgrange}
Scatterplot of the average interaction range $R$ of the obtained campaign solutions (Tabs.~\ref{tab:bandwidth} and~\ref{tab:cutwidth}). The x-axis is the $R=R_{bw}$ optimized in the campaign for the primary objective of the bandwidth $B$, the y-axis is the $R=R_{cw}$ optimized in the campaign for the cutwidth $C$. All points lie below the line $R_{bw}=R_{cw}$, indicating that the $C$ campaign was more effective at lowering $R$.
}
\end{figure}

\begin{figure}[b]
\includegraphics[width=\columnwidth]{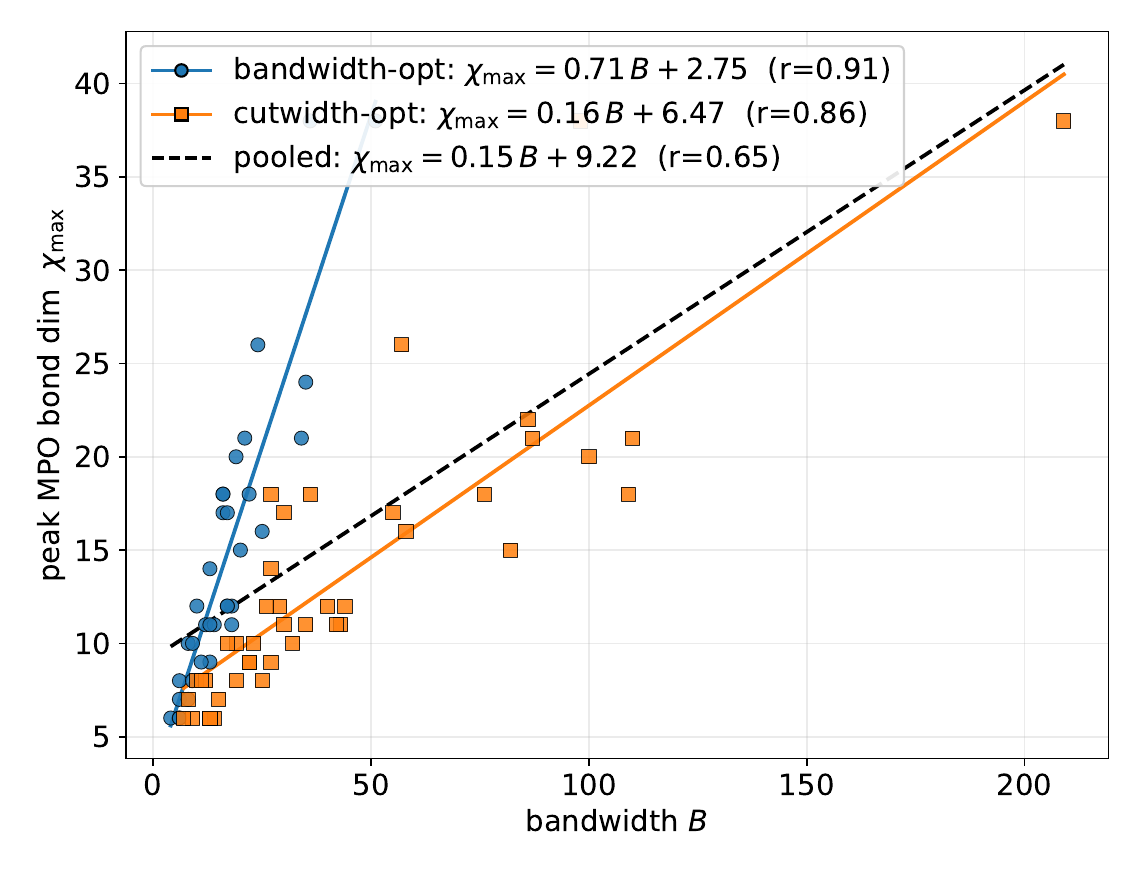}
\caption{\label{fig:mpo_bandwidth}
Peak MPO bond dimension $\chi=\chi_{\text{max}}$ as a function of the bandwidth $B$ does not follow a clear linear law when the data is pooled between the campaigns for $B$ and the cutwidth $C$, though it follows a linear law when the data is fitted separately.
}
\end{figure}

\begin{figure*}[t]
\includegraphics[width=\textwidth]{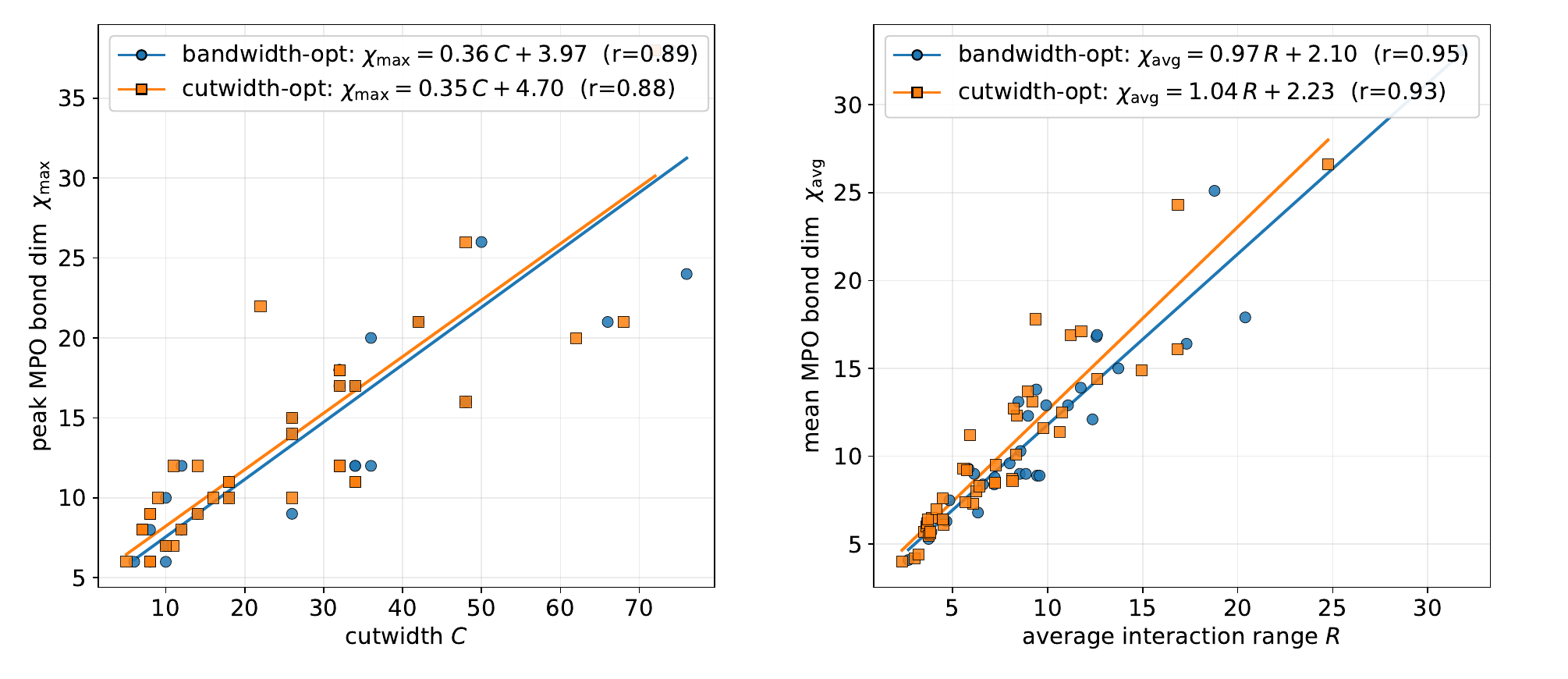}
\caption{\label{fig:mpo_bonddim}
Left: Peak MPO bond dimension $\chi=\chi_{\text{max}}$ as a function of the cutwidth $C$ fitted by a linear law.
Right: Average MPO bond dimension $\chi_{\text{avg}}$ as a function of the average interaction range $R$ is fitted by a linear law.
}
\end{figure*}

\subsection{SAT campaign results}

\subsubsection{\label{sec:certif}Certification}

The central results of the SAT campaigns are shown in Tab.~\ref{tab:bandwidth} for the bandwidth and Tab.~\ref{tab:cutwidth} for the cutwidth.

For $B$, the smallest clusters ($L\lesssim 30$--$40$) are \textit{xsat}-certified; CP-SAT certification extends to $L\sim 100$ for tori and three-dimensional lattices, and to $L\sim 300$ for cylinders, whose quasi-one-dimensional structure makes the bandwidth decision easy. Where the window does not close, it is at least narrow: 1-3 units in all cases except for hyperkagome-324. 

For $C$, thanks to the dedicated phase 2b, we are able to achieve certification for clusters up to $L\sim 100$ for tori, $L\sim 70$ for 3D lattices and all cylinders, though some clusters, including relatively small ones remain open. The windows are 2 units wide except for the hardest clusters hyperkagome-324, pyrochlore-108, pyrochlore-128 and trillium-108.
For hyperkagome-324, the congestion bound of phase 1 gave a lower bound of 50, and the SAT campaign could not improve it, while the optional phase (cf. App.~\ref{app:opt-phase}) improved it to 55. This was the only case where this phase led to an improvement.

\subsubsection{Distribution of $B$, $C$ and $R$}

The next observation is that the $B$ solutions in many cases also converge to the same optimized $C$ as in the $C$-targeted campaign, consistent with the earlier mentioned inequality Eq.~\eqref{eq:ineqBC}. However, there are notable exceptions where the $C$ is noticeably higher. Figure~\ref{fig:mpo_divergence} is a scatter plot of the resulting $C$ vs. $B$. If they converge to the same $C$, the campaign results are connected by a horizontal grey line. Exceptions involve the largest pyrochlore and trillium clusters (to a lesser extent also C$_{12,20,40,60}$). In particular, for pyrochlore-128 the $C$ campaign produces a 1.12$\times$ smaller $C$. In these cases, we can directly test the effect of optimizing for $B$ or for $C$ on the energy. Interestingly, the square torus on $L=100$ sites also falls into this category.

Vice versa, targeting $C$ can greatly increase $B$, e.g. $B=209$ in the hyperkagome-324 case.

Another observation is that optimizing for $C$, we typically get a lot better $R$ as the secondary objective as well. This is visualized in Fig.~\ref{fig:mpo_avgrange} with a scatter plot of $R_{cw}$ from the cutwidth-optimal solutions vs. $R_{bw}$ from the bandwidth-optimal solutions. The points all lie below the $R_{cw}=R_{bw}$ line, indicating that the cutwidth optimization is able to produce smaller $R$ in all cases.

The two observations are likely related: It seems that one can elongate certain bonds to reduce local pileup, trading increased $B$ for reduced $R$.

\subsubsection{MPO bond dimensions}

We can look at how $B$, $C$ and $R$ of the solutions correlate with the MPO bond dimensions. These are highly unintuitive quantities, since the construction is complex and involves a compression of terms. Here the bipartite graph approach from Ref.~\cite{Ren2020MPO} is employed and checked against the MPO compression algorithm from Ref.~\cite{Hubig2017GenericMPO}. The two approaches match in the end result, but the bipartite graph method tends to be faster.
Only the SU(2)-invariant Heisenberg model is employed, and SU(2) symmetry is fully exploited in the code.

Figure~\ref{fig:mpo_bonddim} shows linear fits of the (peak) MPO bond dimension $\chi=\chi_{\text{max}}$ as a function of $C$, and of the average bond dimension $\chi_{\text{avg}}$ as a function of the average interaction range $R$. Both are acceptable linear fits. Remarkably, we get the rule $\chi_{\text{avg}} \approx R+2$ to good accuracy.

When trying to correlate $B$ with the peak MPO $\chi$ (Fig.~\ref{fig:mpo_bandwidth}), however, no clear picture is found. As we have seen, the bandwidth-optimal points have nearly the same $C$, and so the linear law for the blue points in Fig.~\ref{fig:mpo_bandwidth} is essentially inherited from Fig.~\ref{fig:mpo_bonddim}. But the cutwidth-optimal solution may choose to increase $B$ by a lot, leading to a broad scattering of the orange points in Fig.~\ref{fig:mpo_bandwidth}. 

These results already suggest that optimizing for ($C$, $R$) rather than ($B$, $R$) is the better approach. We can further test this by directly comparing energies.

\subsubsection{Energies}

\begin{figure*}
\includegraphics[width=0.7\textwidth]{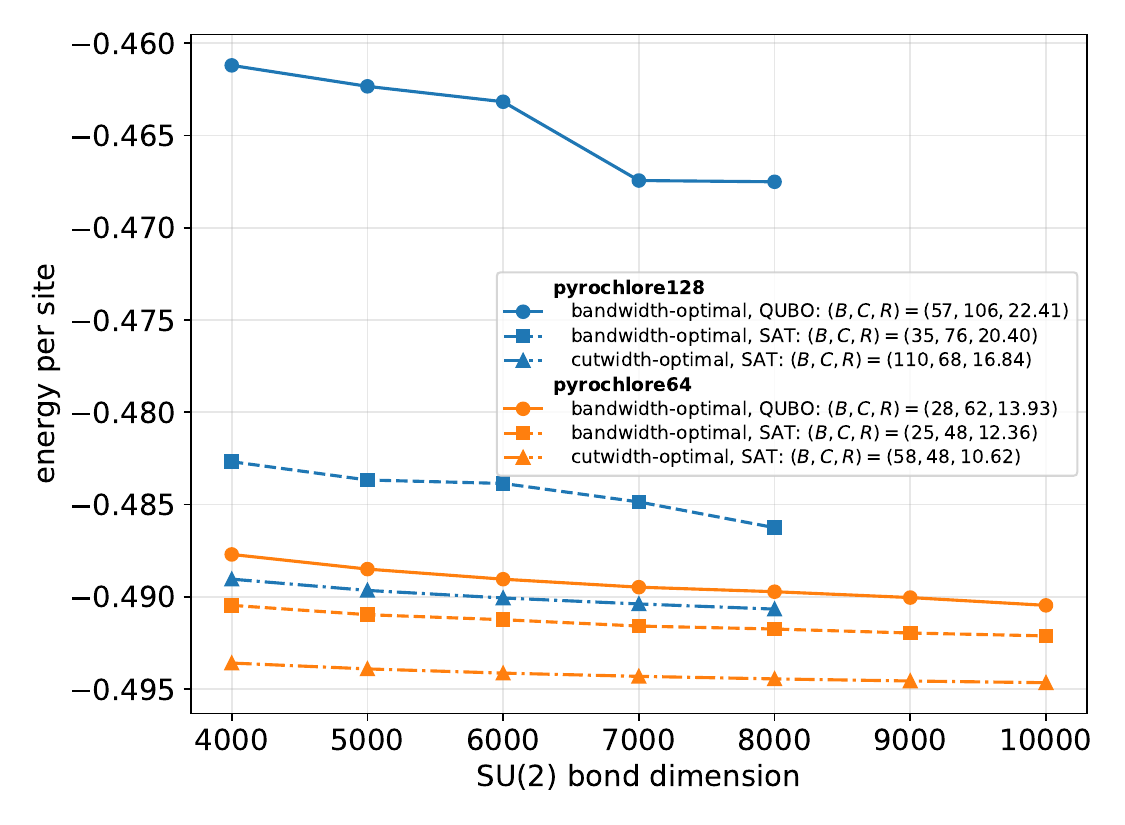}
\caption{\label{fig:E_pyrochlore}
Ground-state energies of the SU(2)-invariant Heisenberg model on the periodic pyrochlore lattice clusters with $L=128$ and $L=64$ (cf. Tabs.~\ref{tab:bandwidth} and~\ref{tab:cutwidth}) as a function of the SU(2)-invariant bond dimension.
}
\end{figure*}

A central question is the dependence of the energy on $B$ and $C$. We can test this for a case where it markedly differs between the campaigns, namely for pyrochlore-128 [$\lr{B,C}=\lr{35,76}$ when $B$ is targeted, $\lr{B,C}=(110,68)$ when $C$ is targeted]. The result is shown in Fig.~\ref{fig:E_pyrochlore}. The difference is quite clear: The bandwidth-optimal result hovers around $E/L=-0.485$, while the cutwidth-optimal one crosses below $E/L=-0.490$. Expressed in terms of bond dimension, DMRG's universal currency, it means that cutwidth-optimal ordering gives better energies at $\chi_{\text{SU(2)}}=4000$, which the bandwidth-optimal ordering cannot achieve even for $\chi_{\text{SU(2)}}=8000$. The QUBO result, which converges to $\lr{B,C}=\lr{57,106}$ is even worse with $E>-0.47$. Clearly, the cutwidth matters much more for the energy.

Figure~\ref{fig:E_pyrochlore} also compares pyrochlore-64, where the campaigns converge to the same $C=48$, but the $C$ campaign achieves a better $R$ ($10.62$ vs. $12.36$). In this case, the $C$ campaign energy is much better as well, and the lines do not cross, i.e. optimizing the ordering buys one more than cranking up the bond dimension.

\begin{figure*}
\includegraphics[width=0.7\textwidth]{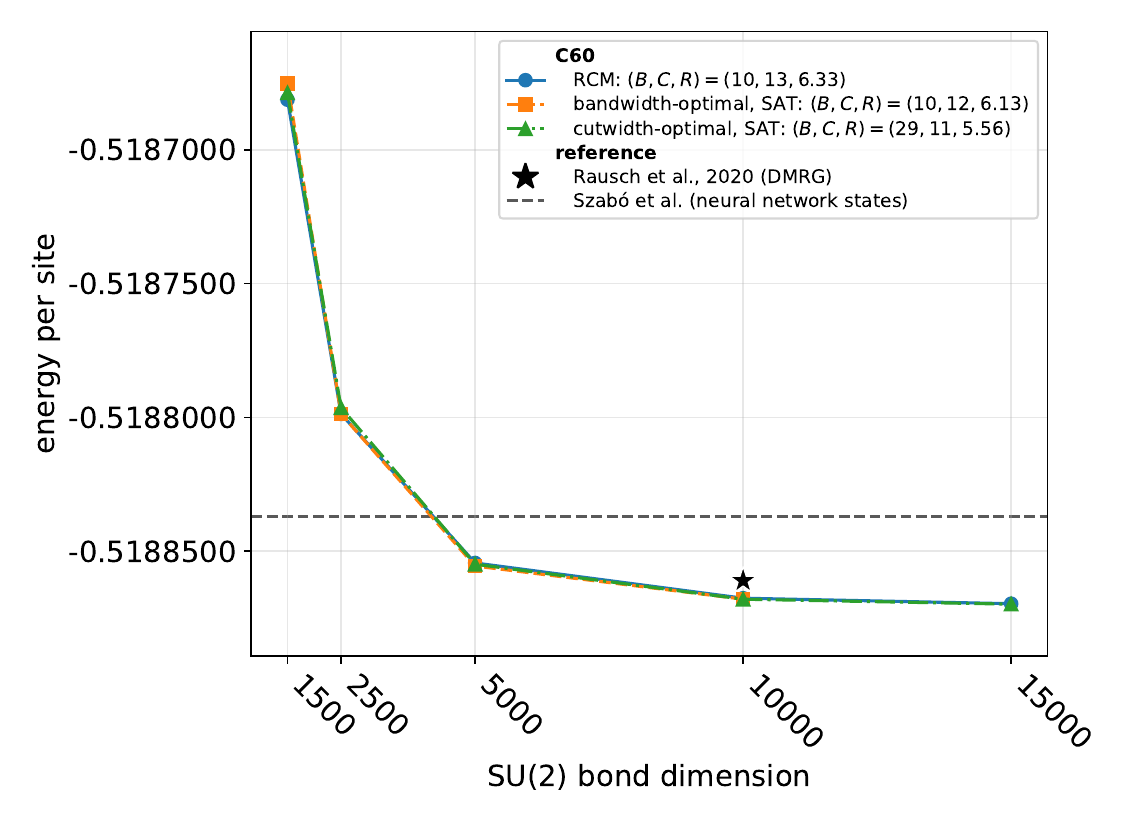}
\caption{\label{fig:E_C60}
Ground-state energies of the SU(2)-invariant Heisenberg model on the C$_{60}$ molecule as a function of the SU(2)-invariant bond dimension. Comparison is with Ref.~\cite{Rausch2021C60} and Ref.~\cite{Szabo2024Fullerene}.
}
\end{figure*}

Finally, we can compare with an easier, but still nontrivial case, namely the Heisenberg model on the C$_{60}$ molecule. The results are shown in Fig.~\ref{fig:E_C60}. Here, both the absolute values and the differences are much smaller: $C=13,12,11$. The resulting energies are a wash, and we can use any enumeration to extrapolate. A highly optimized lattice compiler is therefore mostly of use in the frontier regime of complex lattices, where the MPS is further away from the ground state.

\section{\label{sec:conclusion}Conclusion}

I have discussed the graph bandwidth $B$ (maximum interaction range), cutwidth $C$ (maximum number of cut bonds) and mean edge length $R$ (average interaction range) as proxy metrics to achieve the best enumeration of DMRG sites for complex geometries. The cutwidth is directly correlated with the peak MPO bond dimension, while $R$ is correlated with the average MPO bond dimension. I argue that the best procedure is therefore to optimize over the joined objective ($C$, $R$) lexicographically: First, $C$ is minimized, then $R$ is minimized at fixed $C$. The minimization of $R$ often includes stretching a bond to allow for tighter packing, and may therefore strongly increase the bandwidth $B$. However, $B$ by itself remains largely irrelevant for both MPO bond dimension and final energy quality. The main hurdle for DMRG rather lies in densely packed interactions.

However, bandwidth and cutwidth are correlated in the sense that an enumeration with small $B$ cannot have a large $C$. Hence, minimization of $B$ can in fact be used as a proxy objective to minimize $C$, and may provide a false positive signal that it was the minimization of $B$ that mattered. This route is suboptimal: While there are many breakeven cases, in others I find that targeting $C$ directly can make it significantly smaller; moreover, fixing $B$ hampers the secondary objective of minimizing $R$.

All these statements are of course made based on results obtained numerically as far as the algorithms can carry us. However, the SAT-based approach can put rigorous bounds on the solution beyond just ``best-converged'', and can in fact even certify the best solution for moderately-sized lattice clusters and molecules.

Heuristic approaches, like the Reverse Cuthill--McKee algorithm, can fail to minimize anything even for small problems, and should be rather seen as cheap bound estimates. An interesting open question is whether a cheap, but powerful cutwidth-targeting heuristic can be developed.

Deviations from the optimum do not actually matter much for simple graphs (as quasi-1D systems or small molecules), since we can just throw more bond dimension at the problem.
For the really tough edge cases, however, no amount of bond dimension will be able to buy a drastically improved energy if the enumeration was chosen poorly, and so it pays off to thoroughly invest in a well-optimized enumeration first.

Pathways to extend this investigation include: Mapping the lattice to a ladder, i.e. to blocked-up sites. This is possible in the code, but has not been evaluated in detail. For multiple graphs that correspond to longer-ranged interactions $J_1$, $J_2$, $J_3$, \ldots, the natural objective for the cutwidth campaign is to weigh each bond of the cut by the corresponding $J_n$. If minimizing the bandwidth is really of interest for the multi-graph case, one can do it in a lexicographic way, i.e. first minimize $B_1$ of the $J_1$ graph, then keep it constant (or allow it to be relaxed only slightly) and minimize $B_2$ of the $J_2$ graph; and so on. This feature is also possible in the code. Another approach is to make the enumeration dynamic, i.e. first pre-converge a ground state, and then rearrange, e.g. based on the graph of the spin-spin correlations $C_{ij}=\avg{\mathbf{S}_i\cdot\mathbf{S}_j}$\footnote{
More precisely, perhaps rather based on the concurrence $W_{ij} = \max\lr{0, -2C_{ij} - \frac{1}{2}}$, which cuts off all classical correlations $-\frac{1}{4} \leq C_{ij} \leq \frac{1}{4}$.
}.
Finally, it would be interesting to see whether some of the conclusions change when going to fermions, and one could compare to the genetic algorithm approach~\cite{Zhai2023Block2}.

\section*{Acknowledgments}

Discussions with Matthias Peschke and Sukhbinder Singh are gratefully acknowledged.

\begin{table*}[!hp]
\centering
\caption{%
\label{tab:bandwidth}
Results of the bandwidth optimization: bandwidth $B$, cutwidth $C$, and average interaction range $R$. $L$: number of vertices, $E$: number of edges. 
RCM: reverse Cuthill--McKee.
Best values within each system are shown in bold. 
\textit{xsat}: decisive infeasibility cross-checked by CaDiCaL and Glucose; 
\textit{cpsat}: proven by CP-SAT; otherwise $[\mathrm{LB},\mathrm{UB}]$ is the certified window.}
\renewcommand{\arraystretch}{1.15}
\resizebox{\textwidth}{!}{%
\begin{tabular}{l c c c | c c c | c c c | c c c l}
\toprule
cluster & $L$ & lattice info & $E$ & \multicolumn{3}{c|}{RCM} & \multicolumn{3}{c|}{QUBO} & \multicolumn{3}{c}{SAT} & certif. \\
\cmidrule(lr){5-7}\cmidrule(lr){8-10}\cmidrule(lr){11-13}
 & & & & $B$ & $C$ & $R$ & $B$ & $C$ & $R$ & $B$ & $C$ & $R$ & \\
\midrule
trunc. tetrah. (C$_{12}$)    & 12 &  & 18 & 5 & \textbf{6} & 2.78 & \textbf{4} & \textbf{6} & \textbf{2.67} & \textbf{4} & \textbf{6} & \textbf{2.67} & xsat \\
dodecah. (C$_{20}$)          & 20 &  & 30 & \textbf{6} & \textbf{8} & \textbf{3.73} & \textbf{6} & \textbf{8} & \textbf{3.73} & \textbf{6} & \textbf{8} & \textbf{3.73} & xsat \\
icosidodecah.           & 30 &  & 60 & 10 & 14 & 4.90 & \textbf{9} & \textbf{12} & 4.83 & \textbf{9} & \textbf{12} & \textbf{4.67} & xsat \\
C$_{40}$ fullerene         & 40 &  & 60 & 10 & 13 & 5.60 & 10 & 13 & 5.60 & \textbf{8} & \textbf{10} & \textbf{4.83} & xsat \\
trunc. icosah. (C$_{60}$)    & 60 &  & 90 & \textbf{10} & 13 & 6.33 & \textbf{10} & 13 & 6.33 & \textbf{10} & \textbf{12} & \textbf{6.13} & cpsat \\
\midrule
kagome Y cylinder       & 192 & $16\times8$ & 376 & 10 & 12 & 4.65 & \textbf{8} & \textbf{10} & 4.55 & \textbf{8} & \textbf{10} & \textbf{4.52} & cpsat \\
kagome Y cylinder       & 288 & $16\times12$ & 564 & 15 & 18 & 6.82 & 12 & \textbf{14} & \textbf{6.56} & \textbf{11} & \textbf{14} & 6.59 & cpsat \\
kagome Y cylinder       & 344 & $7\times4$ full stars & 672 & 20 & 24 & 9.19 &  &  &  & \textbf{14} & \textbf{18} & \textbf{8.57} & cpsat \\
kagome torus            & 48 & $4\times4$ & 96 & 16 & 22 & 7.75 & 16 & 22 & 7.75 & \textbf{13} & \textbf{18} & \textbf{7.21} & cpsat \\
kagome torus            & 108 & $6\times6$ & 216 & 26 & 34 & 11.94 & 26 & 34 & 11.94 & \textbf{20} & \textbf{26} & \textbf{11.07} & $[18,20]$ \\
\midrule
triangular Y cylinder   & 128 & $16\times8$ & 368 & 12 & 22 & 6.02 & \textbf{9} & \textbf{18} & \textbf{5.83} & \textbf{9} & \textbf{18} & \textbf{5.83} & cpsat \\
triangular Y cylinder   & 192 & $16\times12$ & 552 & 18 & 34 & 9.06 & \textbf{13} & \textbf{26} & \textbf{8.46} & \textbf{13} & \textbf{26} & \textbf{8.46} & cpsat \\
triangular torus        & 64 & $8\times8$ & 192 & 22 & 44 & 10.33 & 22 & 44 & 10.33 & \textbf{17} & \textbf{34} & \textbf{9.92} & cpsat \\
triangular torus        & 100 & $10\times10$ & 300 & 28 & 56 & 13.08 & 28 & 56 & 13.08 & \textbf{21} & \textbf{42} & \textbf{12.60} & cpsat \\
\midrule
square torus            & 100 & $10\times10$ & 200 & \textbf{19} & \textbf{36} & 12.62 &  &  &  & \textbf{19} & \textbf{36} & \textbf{11.74} & $[16,19]$ \\
\midrule
hyperkagome             & 96 & $2\times2\times2$ & 192 & 39 & 52 & 15.72 & 37 & 54 & 16.60 & \textbf{22} & \textbf{32} & \textbf{13.72} & cpsat \\
hyperkagome             & 324 & $3\times3\times3$ & 648 & 87 & 116 & 37.19 & 84 & 116 & 37.73 & \textbf{51} & \textbf{72} & \textbf{31.82} & $[44,51]$ \\
\midrule
pyrochlore              & 32 & $2\times2\times2$ & 96 & 16 & 28 & \textbf{6.25} & 15 & 32 & 7.38 & \textbf{13} & \textbf{26} & 6.33 & xsat \\
pyrochlore              & 48 & tilted & 144 & 28 & 46 & 9.56 & 22 & 42 & 10.35 & \textbf{18} & \textbf{34} & \textbf{9.44} & cpsat \\
pyrochlore              & 48 & tilted & 144 & 32 & 52 & 10.62 & 22 & 42 & 10.36 & \textbf{18} & \textbf{36} & \textbf{9.57} & cpsat \\
pyrochlore              & 48 & tilted & 144 & 26 & 44 & 9.39 & 19 & 38 & 9.31 & \textbf{17} & \textbf{34} & \textbf{8.53} & cpsat \\
pyrochlore              & 48 & tilted & 144 & 26 & 46 & 9.58 & 21 & 42 & 9.71 & \textbf{17} & \textbf{34} & \textbf{8.85} & cpsat \\
pyrochlore              & 64 & tilted & 192 & 36 & 60 & 13.15 & 28 & 62 & 13.93 & \textbf{25} & \textbf{48} & \textbf{12.36} & $[22,25]$ \\
pyrochlore              & 108 & $3\times3\times3$ & 324 & 48 & 80 & 17.88 & 38 & 80 & 18.48 & \textbf{34} & \textbf{66} & \textbf{17.31} & $[32,34]$ \\
pyrochlore              & 128 & tilted & 384 & 60 & 104 & 21.11 & 57 & 106 & 22.41 & \textbf{35} & \textbf{76} & \textbf{20.40} & $[32,35]$ \\
\midrule
trillium                & 32 & $2\times2\times2$ & 96 & 22 & 40 & 9.00 & 17 & 44 & 9.29 & \textbf{16} & \textbf{32} & \textbf{8.00} & cpsat \\
trillium                & 48 & $3\times2\times2$ & 144 & 27 & 44 & 10.32 & 20 & 38 & 9.39 & \textbf{16} & \textbf{32} & \textbf{8.97} & cpsat \\
trillium                & 64 & $4\times2\times2$ & 192 & 27 & 44 & 10.51 & 22 & 44 & 10.51 & \textbf{16} & \textbf{32} & \textbf{9.40} & cpsat \\
trillium                & 72 & $3\times3\times2$ & 216 & 32 & 62 & 14.44 & 31 & 60 & 14.24 & \textbf{24} & \textbf{50} & \textbf{12.57} & $[23,24]$ \\
trillium                & 108 & $3\times3\times3$ & 324 & 52 & 98 & 21.97 & 48 & 98 & 23.31 & \textbf{36} & \textbf{74} & \textbf{18.78} & $[33,36]$ \\
\bottomrule
\end{tabular}%
}
\end{table*}


\begin{table*}[!hp]
\centering
\caption{%
\label{tab:cutwidth}
Results of the cutwidth optimization: bandwidth $B$, cutwidth $C$, and average interaction range $R$.
$L$: number of vertices, $E$: number of edges. RCM is the reverse Cuthill--McKee baseline; SAT is the cutwidth-optimized ordering. 
Best values within each system are shown in bold. 
\textit{cpsat}: optimal cutwidth proven by CP-SAT. 
\textit{xsat}: optimal cutwidth proven by the subset proofs of phase 2b, confirmed by two independent SAT solvers. 
Otherwise $[\mathrm{LB},\mathrm{UB}]$ is the certified window.}
\renewcommand{\arraystretch}{1.1}
\resizebox{\textwidth}{!}{%
\begin{tabular}{l c c c | c c c | c c c l}
\toprule
cluster & $L$ & lattice info & $E$ & \multicolumn{3}{c|}{RCM} & \multicolumn{3}{c}{SAT (cutwidth)} & certif. \\
\cmidrule(lr){5-7}\cmidrule(lr){8-10}
 & & & & $B$ & $C$ & $R$ & $B$ & $C$ & $R$ & \\
\midrule
trunc. tetrah. (C$_{12}$)    & 12 &  & 18 & \textbf{5} & 6 & 2.78 & 9 & \textbf{5} & \textbf{2.33} & xsat \\
dodecah. (C$_{20}$)          & 20 &  & 30 & \textbf{6} & 8 & 3.73 & 10 & \textbf{7} & \textbf{3.47} & xsat \\
icosidodecah.           & 30 &  & 60 & \textbf{10} & 14 & 4.90 & 19 & \textbf{12} & \textbf{4.53} & xsat \\
C$_{40}$ fullerene         & 40 &  & 60 & \textbf{10} & 13 & 5.60 & 19 & \textbf{9} & \textbf{4.47} & xsat \\
trunc. icosah. (C$_{60}$)    & 60 &  & 90 & \textbf{10} & 13 & 6.33 & 29 & \textbf{11} & \textbf{5.56} & xsat \\
\midrule
kagome Y cylinder       & 192 & $16\times8$ & 376 & \textbf{10} & 12 & 4.65 & 15 & \textbf{10} & \textbf{4.48} & xsat \\
kagome Y cylinder       & 288 & $16\times12$ & 564 & \textbf{15} & 18 & 6.82 & 22 & \textbf{14} & \textbf{6.41} & xsat \\
kagome Y cylinder       & 344 & $7\times4$ full stars & 672 & \textbf{20} & 24 & 9.19 & 30 & \textbf{18} & \textbf{8.32} & xsat \\
kagome torus            & 48 & $4\times4$ & 96 & \textbf{16} & 22 & 7.75 & 43 & \textbf{18} & \textbf{6.25} & $[16,18]$ \\
kagome torus            & 108 & $6\times6$ & 216 & \textbf{26} & 34 & 11.94 & 82 & \textbf{26} & \textbf{9.77} & xsat \\
\midrule
triangular Y cylinder   & 128 & $16\times8$ & 368 & \textbf{12} & 22 & 6.02 & 17 & \textbf{18} & \textbf{5.73} & xsat \\
triangular Y cylinder   & 192 & $16\times12$ & 552 & \textbf{18} & 34 & 9.06 & 27 & \textbf{26} & \textbf{8.21} & xsat \\
triangular torus        & 64 & $8\times8$ & 192 & \textbf{22} & 44 & 10.33 & 55 & \textbf{34} & \textbf{9.19} & xsat \\
triangular torus        & 100 & $10\times10$ & 300 & \textbf{28} & 56 & 13.08 & 87 & \textbf{42} & \textbf{11.76} & xsat \\
\midrule
square torus            & 100 & $10\times10$ & 200 & \textbf{19} & 36 & 12.62 & 86 & \textbf{22} & \textbf{9.36} & xsat \\
\midrule
hyperkagome             & 96 & $2\times2\times2$ & 192 & \textbf{39} & 52 & 15.72 & 76 & \textbf{32} & \textbf{10.74} & $[30,32]$ \\
hyperkagome             & 324 & $3\times3\times3$ & 648 & \textbf{87} & 116 & 37.19 & 209 & \textbf{72} & \textbf{24.75} & $[55,72]$ \\
\midrule
pyrochlore              & 32 & $2\times2\times2$ & 96 & \textbf{16} & 28 & 6.25 & 23 & \textbf{26} & \textbf{6.08} & $[24,26]$ \\
pyrochlore              & 48 & tilted & 144 & \textbf{28} & 46 & 9.56 & 42 & \textbf{34} & \textbf{8.11} & xsat \\
pyrochlore              & 48 & tilted & 144 & \textbf{32} & 52 & 10.62 & 35 & \textbf{34} & \textbf{8.15} & xsat \\
pyrochlore              & 48 & tilted & 144 & \textbf{26} & 44 & 9.39 & 40 & \textbf{32} & \textbf{7.22} & $[30,32]$ \\
pyrochlore              & 48 & tilted & 144 & \textbf{26} & 46 & 9.58 & 44 & \textbf{32} & \textbf{7.22} & $[30,32]$ \\
pyrochlore              & 64 & tilted & 192 & \textbf{36} & 60 & 13.15 & 58 & \textbf{48} & \textbf{10.62} & $[46,48]$ \\
pyrochlore              & 108 & $3\times3\times3$ & 324 & \textbf{48} & 80 & 17.88 & 100 & \textbf{62} & \textbf{14.95} & $[52,62]$ \\
pyrochlore              & 128 & tilted & 384 & \textbf{60} & 104 & 21.11 & 110 & \textbf{68} & \textbf{16.84} & $[60,68]$ \\
\midrule
trillium                & 32 & $2\times2\times2$ & 96 & \textbf{22} & 40 & 9.00 & 30 & \textbf{32} & \textbf{7.27} & xsat \\
trillium                & 48 & $3\times2\times2$ & 144 & \textbf{27} & 44 & 10.32 & \textbf{27} & \textbf{32} & \textbf{8.39} & xsat \\
trillium                & 64 & $4\times2\times2$ & 192 & \textbf{27} & 44 & 10.51 & 36 & \textbf{32} & \textbf{8.96} & xsat \\
trillium                & 72 & $3\times3\times2$ & 216 & \textbf{32} & 62 & 14.44 & 57 & \textbf{48} & \textbf{11.22} & xsat \\
trillium                & 108 & $3\times3\times3$ & 324 & \textbf{52} & 98 & 21.97 & 98 & \textbf{72} & \textbf{16.85} & $[56,72]$ \\
\bottomrule
\end{tabular}%
}
\end{table*}


\clearpage

\bibliographystyle{plainurl}
\bibliography{references,sat_routing}

\clearpage

\appendix
\section{\label{app:heuristic-benchmark}Heuristic algorithms benchmark}

Table~\ref{tab:heuristic} shows the results of the heuristic algorithms for the given clusters.

\begin{table*}[p]
\centering
\caption{%
Bandwidth $B$, cutwidth $C$, and average interaction range $R$ for various
cluster geometries and ordering algorithms. $L$: number of vertices;
$E$: number of edges. 
RCM: Reverse Cuthill-McKee, GPS: Gibbs--Poole--Stockmeyer. 
The ``manual'' enumeration was set by hand or is returned by the lattice generation algorithm, and is not necessarily optimized.
Best values within each system are shown in bold.
RCM is SciPy's \texttt{reverse\_cuthill\_mckee} reference implementation
(the Boost \texttt{cuthill\_mckee\_ordering} used by the {\Cpp} code yields the
same metrics).
}
\label{tab:heuristic}
\renewcommand{\arraystretch}{1.15}

\resizebox{\textwidth}{!}{%
\begin{tabular}{l c c c | ccc | ccc | ccc | ccc | ccc | ccc}
\toprule
cluster & $L$ & lattice info & $E$
& \multicolumn{3}{c|}{Manual}
& \multicolumn{3}{c|}{RCM}
& \multicolumn{3}{c|}{GPS}
& \multicolumn{3}{c|}{Sloan}
& \multicolumn{3}{c|}{King}
& \multicolumn{3}{c}{Spectral} \\
\cmidrule(lr){5-7}
\cmidrule(lr){8-10}
\cmidrule(lr){11-13}
\cmidrule(lr){14-16}
\cmidrule(lr){17-19}
\cmidrule(lr){20-22}
& & & &
$B$ & $C$ & $R$ &
$B$ & $C$ & $R$ &
$B$ & $C$ & $R$ &
$B$ & $C$ & $R$ &
$B$ & $C$ & $R$ &
$B$ & $C$ & $R$ \\
\midrule

trunc. tetrah. (C$_{12}$)
& 12 &  & 18
& \textbf{5} & 6 & 2.78
& \textbf{5} & 6 & 2.78
& 6 & 7 & 3.00
& 8 & \textbf{5} & \textbf{2.44}
& 9 & 6 & \textbf{2.44}
& \textbf{5} & 6 & 2.67 \\

dodecah. (C$_{20}$)
& 20 &  & 30
& 10 & \textbf{7} & \textbf{3.47}
& \textbf{6} & 8 & 3.73
& 8 & \textbf{7} & 3.60
& 9 & 8 & 3.67
& 17 & 9 & 3.87
& 7 & 8 & 3.73 \\

icosidodecah.
& 30 &  & 60
& 28 & \textbf{12} & 5.23
& \textbf{10} & 14 & 4.90
& 13 & 20 & 5.90
& 14 & 14 & 4.83
& 23 & 16 & 5.20
& \textbf{10} & \textbf{12} & \textbf{4.73} \\

C$_{40}$ fullerene
& 40 &  & 60
& 16 & \textbf{11} & \textbf{5.03}
& \textbf{10} & 13 & 5.60
& 16 & 15 & 5.83
& 18 & 13 & 5.60
& 34 & 12 & 5.30
& \textbf{10} & 12 & 5.13 \\

trunc. icosah. (C$_{60}$)
& 60 &  & 90
& 15 & \textbf{11} & \textbf{5.69}
& \textbf{10} & 13 & 6.33
& 18 & 15 & 6.93
& 21 & 13 & 6.04
& 54 & 14 & 6.38
& 14 & 14 & 6.20 \\

\midrule

kagome Y cylinder
& 192 & $16\times8$ & 376
& \textbf{8} & \textbf{10} & \textbf{4.55}
& 10 & 12 & 4.65
& 14 & 12 & 4.68
& 15 & \textbf{10} & 4.66
& 185 & 68 & 18.89
& 11 & 14 & 4.74 \\

kagome Y cylinder
& 288 & $16\times12$ & 564
& \textbf{12} & \textbf{14} & \textbf{6.56}
& 15 & 18 & 6.82
& 21 & 18 & 6.88
& 24 & 16 & 6.72
& 277 & 76 & 22.35
& 17 & 18 & 6.88 \\

kagome Y cylinder
& 344 & $7\times4$ full stars & 672
& \textbf{16} & \textbf{18} & \textbf{8.63}
& 20 & 24 & 9.19
& 28 & 24 & 9.30
& 42 & 22 & 8.77
& 342 & 66 & 20.15
& 21 & 24 & 8.85 \\

kagome torus
& 48 & $4\times4$ & 96
& 40 & \textbf{18} & 7.08
& \textbf{16} & 22 & 7.75
& 25 & 28 & 9.02
& 34 & 22 & 7.19
& 40 & \textbf{18} & \textbf{6.92}
& 23 & 22 & 7.31 \\

kagome torus
& 108 & $6\times6$ & 216
& 94 & \textbf{26} & 11.06
& \textbf{26} & 34 & 11.94
& 42 & 40 & 13.61
& 65 & 34 & 11.04
& 90 & 28 & \textbf{10.99}
& 37 & 42 & 12.60 \\

\midrule

triangular Y cylinder
& 128 & $16\times8$ & 368
& \textbf{9} & \textbf{18} & 5.83
& 12 & 22 & 6.02
& 19 & 32 & 6.90
& 15 & \textbf{18} & \textbf{5.78}
& 23 & 20 & 5.81
& 15 & 22 & 6.03 \\

triangular Y cylinder
& 192 & $16\times12$ & 552
& \textbf{13} & \textbf{26} & 8.46
& 18 & 34 & 9.06
& 31 & 44 & 10.67
& 23 & 30 & 8.47
& 35 & 28 & \textbf{8.36}
& 23 & 32 & 8.76 \\

triangular torus
& 64 & $8\times8$ & 192
& 57 & \textbf{34} & 9.92
& \textbf{22} & 44 & 10.33
& 34 & 54 & 11.67
& 35 & 42 & 9.28
& 59 & 36 & \textbf{8.93}
& 28 & 44 & 10.31 \\

triangular torus
& 100 & $10\times10$ & 300
& 91 & \textbf{42} & 12.60
& \textbf{28} & 56 & 13.08
& 44 & 70 & 14.97
& 48 & 56 & 11.85
& 93 & 44 & \textbf{11.31}
& 45 & 56 & 12.30 \\

\midrule

square torus
& 100 & $10\times10$ & 200
& 90 & \textbf{22} & \textbf{9.90}
& \textbf{19} & 36 & 12.62
& 31 & 36 & 13.82
& 22 & 36 & 12.59
& 76 & 38 & 12.57
& 27 & 32 & 11.18 \\

\midrule

hyperkagome
& 96 & $2\times2\times2$ & 192
& 74 & 44 & 14.71
& \textbf{39} & 52 & 15.72
& 59 & 68 & 19.46
& 67 & \textbf{36} & \textbf{12.00}
& 64 & \textbf{36} & 12.30
& \textbf{39} & \textbf{36} & 12.25 \\

hyperkagome
& 324 & $3\times3\times3$ & 648
& 286 & 104 & 34.14
& 87 & 116 & 37.19
& 145 & 162 & 47.37
& 194 & \textbf{84} & 29.56
& 285 & 86 & 29.66
& \textbf{85} & \textbf{84} & \textbf{28.18} \\

\midrule

pyrochlore
& 32 & $2\times2\times2$ & 96
& 25 & 30 & 6.50
& \textbf{16} & 28 & \textbf{6.25}
& 21 & 42 & 8.04
& 23 & 28 & 6.40
& 31 & 28 & 6.67
& \textbf{16} & \textbf{26} & 6.42 \\

pyrochlore
& 48 & tilted (a) & 144
& 41 & 42 & 9.08
& \textbf{28} & 46 & 9.56
& 35 & 58 & 11.62
& 38 & \textbf{40} & \textbf{9.07}
& 36 & \textbf{40} & 9.12
& 29 & \textbf{40} & 9.51 \\

pyrochlore
& 48 & tilted (b) & 144
& 42 & 44 & 9.25
& 32 & 52 & 10.62
& 38 & 54 & 11.06
& 40 & \textbf{38} & 9.07
& 44 & \textbf{38} & \textbf{9.03}
& \textbf{26} & 40 & 9.36 \\

pyrochlore
& 48 & tilted (c) & 144
& 45 & 40 & 9.11
& 26 & 44 & 9.39
& 34 & 58 & 11.57
& 37 & 40 & 9.12
& 41 & 40 & 9.11
& \textbf{25} & \textbf{34} & \textbf{8.32} \\

pyrochlore
& 48 & tilted (d) & 144
& 45 & 44 & 9.35
& \textbf{26} & 46 & 9.58
& 35 & 54 & 10.92
& 35 & 40 & 9.19
& 41 & 42 & 9.08
& 28 & \textbf{32} & \textbf{7.60} \\

pyrochlore
& 64 & tilted & 192
& 57 & 54 & 11.93
& \textbf{36} & 60 & 13.15
& 46 & 90 & 16.83
& 51 & \textbf{52} & 11.62
& 53 & \textbf{52} & \textbf{11.45}
& 40 & \textbf{52} & 12.65 \\

pyrochlore
& 108 & $3\times3\times3$ & 324
& 83 & 72 & 15.89
& \textbf{48} & 80 & 17.88
& 69 & 126 & 23.11
& 92 & 70 & 15.71
& 83 & \textbf{68} & \textbf{15.69}
& 55 & 72 & 16.88 \\

pyrochlore
& 128 & tilted & 384
& 113 & 106 & 22.31
& 60 & 104 & 21.11
& 87 & 138 & 27.32
& 100 & \textbf{76} & \textbf{18.14}
& 112 & 82 & 18.42
& \textbf{59} & \textbf{76} & 19.31 \\

\midrule

trillium
& 32 & $2\times2\times2$ & 96
& 26 & 38 & 8.33
& \textbf{22} & 40 & 9.00
& 25 & 46 & 9.77
& 27 & 42 & 8.69
& 29 & 38 & 8.27
& \textbf{22} & \textbf{32} & \textbf{8.04} \\

trillium
& 48 & $3\times2\times2$ & 144
& 42 & 48 & 10.00
& 27 & 44 & 10.32
& 31 & 56 & 12.15
& 33 & 40 & \textbf{9.24}
& 43 & 40 & 9.40
& \textbf{19} & \textbf{32} & 9.33 \\

trillium
& 64 & $4\times2\times2$ & 192
& 58 & 48 & 10.83
& 27 & 44 & 10.51
& 36 & 62 & 12.83
& 26 & 40 & 9.66
& 41 & 36 & \textbf{9.46}
& \textbf{19} & \textbf{32} & 9.67 \\

trillium
& 72 & $3\times3\times2$ & 216
& 66 & 64 & 14.00
& \textbf{32} & 62 & 14.44
& 48 & 78 & 16.53
& 55 & 56 & 13.06
& 56 & 58 & 12.65
& 38 & \textbf{52} & \textbf{12.19} \\

trillium
& 108 & $3\times3\times3$ & 324
& 98 & 94 & 20.89
& \textbf{52} & 98 & 21.97
& 71 & 122 & 25.70
& 76 & 92 & 20.17
& 97 & 94 & 19.81
& 58 & \textbf{80} & \textbf{18.31} \\

\bottomrule
\end{tabular}%
}
\end{table*}

\section{\label{app:qubo}QUBO approach}

Reference~\cite{GuoDinneen2025} proposes different mappings of the linear bandwidth problem to a QUBO. Briefly, one introduces binary assignment variables
\[
x_{v,i}\in\{0,1\},\qquad v\in V,\quad i=0,\ldots,n-1,
\]
where \(x_{v,i}=1\) means that vertex \(v\) is placed at position \(i\).  All
formulations include the standard permutation penalty
\begin{equation}
\begin{split}
P_{\mathrm{perm}} &= A\sum_{v\in V}\left(\sum_i x_{v,i}-1\right)^2+\\
&A\sum_i\left(\sum_{v\in V}x_{v,i}-1\right)^2 ,
\end{split}
\end{equation}
which enforces that each vertex occupies exactly one position and each position
is occupied by exactly one vertex.

The main approach uses a decision-based QUBO.  For a trial bandwidth $B$, one penalizes
all edge assignments whose endpoint positions differ by more than $B$:
\[
P_B
=
\beta \sum_{(u,v)\in E}
\sum_{\substack{i,j\\ |i-j|>B}}
x_{u,i}x_{v,j}.
\]
The QUBO objective function
\[
Q_k(x)=P_{\mathrm{perm}}+P_B
\]
has low-energy feasible states corresponding to permutations with bandwidth at most $B$.  One scans over candidate values of $B$, starting from a simple lower bound, solves the corresponding QUBO instances and
retains the best feasible permutation found.

An alternative approach is the exponential-penalty formulation in which the fixed threshold $B$ is
replaced by a distance-dependent edge penalty,
\[
Q_{\exp}(x)
=
P_{\mathrm{perm}}
+
B\sum_{(u,v)\in E}
\sum_{i,j}
\alpha^{|i-j|}x_{u,i}x_{v,j},
\]
with optional zeroing out values for large distances. This gives a single QUBO solve that directly favors small $B$, at the cost of larger coefficient ranges. 

Having implemented both approaches in both {\Cpp} and Python coupled with the tabu solver, I find that the exponential approach is way too costly to be of practical use, and therefore I only use the decision method. Running it for various hyperparameters on selected clusters (see Tab.~\ref{tab:bandwidth}) can in general improve on the RCM result, but I could only match the SAT approach for very small clusters and for ladders. For more complex clusters, the improvement over RCM is very slight despite hours of runtime. The QUBO code is provided in the repository for others to experiment with.

\section{\label{app:opt-phase}Optional phase: improved congestion bound ($C$ only)}

\begin{figure}
\includegraphics[width=\columnwidth]{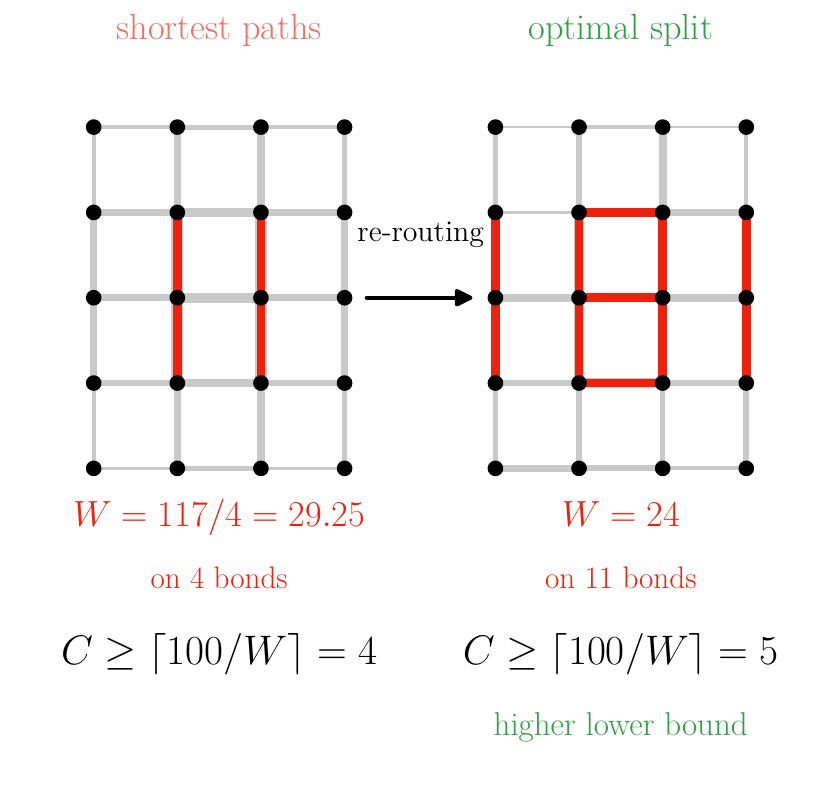}
\caption{\label{fig:certify_congestion_lp}
Visual guide to the improved congestion bound for a $5\times4$ plaquette with open boundary conditions. 
Left: if every message is split evenly among its shortest paths, the busiest bonds (red) carry $W=29.25$ messages, so that $C\ge4$. 
Right: the optimal routing spreads the traffic over more bonds and lowers the maximal load to $W=24$, so that $C\ge5$. The line thickness indicates the load of a bond.
}
\end{figure}

The congestion bound of phase 1 (Sec.~\ref{sec:sat}) can be improved in a separate phase, which is not part of the default campaign, cf. Fig.~\ref{fig:certify_congestion_lp}. The shortest paths are not essential to the argument leading to Eq.~\eqref{eq:congestion}: any way of routing the messages yields a valid bound, and the smaller the load $W$ on the busiest bond, the stronger it is. Shortest paths tend to crowd the traffic onto a few central bonds while others remain underused. One can therefore ask for the best possible routing, in which each message may be split into fractions that travel along different routes, such that $W$ becomes as small as possible. This is a \textit{linear program}, i.e.\ the minimization of a linear quantity ($W$) under linear constraints (the message of every pair is delivered, nothing is lost on the way, and no bond carries more than $W$), a type of problem that standard solvers handle exactly. It is known as the multicommodity flow problem~\cite{Leighton1999}. The result is inserted into Eq.~\eqref{eq:congestion} as before.
 To remain rigorous, the code does not rely on the optimal value reported by the solver, but recomputes $W$ from the routes that it returns, after checking that they deliver every message.

If all bonds of the cluster are equivalent by symmetry, as on the tori and on most clusters considered here, the shortest paths are already optimal and nothing is gained. Otherwise there can be a gain. The linear program has $2EL$ unknowns ($E$: number of bonds) and becomes expensive for large clusters. In the clusters considered, it only led to an improvement of the lower bound for hyperkagome-324, which took several hours on one core. This standalone phase might thus be worth a try if the cutwidth window remains large and the bonds of the cluster are not all symmetry-equivalent.

\end{document}